\documentclass[aps,prb,showpacs,reprint]{revtex4-1}

\usepackage{graphicx}
\usepackage{amsmath}
\usepackage{bbm}

\begin{document}
\title{Coexistence of spin-triplet superconductivity with magnetic ordering in
an
orbitally degenerate system: Hartree-Fock-BCS approximation revisited}
\author{Micha{\l} Zegrodnik}
\affiliation{AGH University of Science and Technology,
Faculty of Physics and Applied Computer Science, Al. Mickiewicza 30,
30-059 Krak\'{o}w, Poland}
\author{Jozef Spa{\l}ek}
\affiliation{Marian Smoluchowski Institute of Physics, 
Jagiellonian University, ul. Reymonta 4,
30-059 Krak\'ow, Poland and\\
AGH University of Science and Technology,
Faculty of Physics and Applied Computer Science, Al. Mickiewicza 30,
30-059 Krak\'{o}w, Poland}

\begin{abstract}
The Hund's-rule-exchange induced and coexisting spin-triplet
paired and magnetic states are considered within the doubly degenerate
Hubbard
model with interband hybridization. The Hartree-Fock approximation combined
with the Bardeen-Cooper-Schrieffer (BCS) approach is analyzed for the
case of square
lattice. The calculated phase diagram contains
regions of stability of the spin-triplet superconducting phase coexisting with
either ferromagnetism or antiferromagnetism, as well as a pure superconducting
phase. The influence of the inter-site hybridization on the
stability of the considered phases, as well as the temperature dependence of
both
the
magnetic moment and the superconducting gaps, are also discussed. Our approach
supplements the well known phase diagrams containing only magnetic
phases with the paired triplet states treated on the same footing. We also discuss briefly how to include
the spin fluctuations within this model with real space pairing.
\end{abstract}

\pacs{74.20.-z, 74.25.Dw, 75.10.Lp}

\maketitle

\section{Introduction}\label{sec:intro}

The spin-triplet superconducting phase is believed to appear in Sr$_2$RuO$_4$
\cite{Maeno1994}, UGe$_2$ \cite{Saxena2000}, and URhGe \cite{Tateiwa2001}. In
the last two compounds the considered type of superconducting
phase occurs as coexisting with ferromagnetism. Additionally, even though U
atoms in this compounds contain 5f electrons responsible for magnetism, this multiple-band
system can be regarded as weakly or moderately correlated electron system,
particularly at higher pressure. Originally it has been suggested via a proper quantitative analysis
\cite{Klejnberg1999}$^-$\cite{Zegrodnik2011},
that the the intra atomic Hund's rule
exchange can lead in a natural manner to the coexistence of
superconductivity with magnetic ordering - ferromagnetism or
antiferromagnetism. 

The coexisting superconducting
and magnetic phases are discussed in this work within an orbitally degenerate two-band
Hubbard model using the Hartree-Fock approximation (HF), here
combined with the Bardeen-Cooper-Schrieffer (BCS) approach, i.e., in the
vicinity of the ferromagnetism disappearance, where also the superconductivity
occurs. The particular emphasis is put on the appearance of superconductivity
near the Stoner threshold, where the Hartree-Fock-BCS approximation can be
regarded as realistic. This type of approach can be formulated also for other
systems \cite{Puetter2003}.

The alternative suggested mechanism
for appearance of superconductivity in those systems is the pairing mediated by ferromagnetic
spin fluctuations, which can also appear in the paramagnetic or weakly ferromagnetic
phase \cite{Anderson}. Here, the mean-field approximation provides not only the starting
magnetic phase diagram but also a related discussion of the
superconducting states treated on equal footing. In this approach the spin-fluctuation contribution appears as a next-order contribution. This is the reason for undertaking a revision of the standard Hartree-Fock approximation. Namely, we concentrate here on the spin triplet states, pure
and coexisting with either ferromagnetism or antiferromagnetism, depending on the relative
magnitude of microscopic parameters: the Hubbard intra- and inter-orbital interactions, $U$ and $U^{\prime}$, respectively, the Hund's rule
ferromagnetic exchange integral $J$, the relative magnitude of hybridization
$\beta_h$, and the band filling $n$. The bare band width $W$ is taken as unit of
energy. In the concluding Section we discuss briefly, how to outline 
the approach to include also the quantum fluctuations around this
HF-BCS (saddle point) state, as a higher-order contribution.

The role of exchange interactions is crucial in both the so-called t-J model of
high temperature superconductivity \cite{Spalek1988(2)} and in the so-called Kondo-mediated
pairing in heavy fermion systems \cite{Spalek1988}. In this and the following papers we discuss the idea of real space pairing for the triplet-paired states in the regime of weakly correlated particles and include both the inter-band hybridization and the corresponding Coulomb interactions. We think
that this relatively simple approach is relevant to the mentioned at the
beginning ferromagnetic superconductors because of the following reasons.
Although the effective exchange (Weiss-type) field acts only on the
spin degrees of freedom, it is important in determining the second critical
field of ferromagnetic superconductor in the so-called Pauli limit \cite{Clagton}, as the
orbital effects in the Cooper-pair breaking process are then negligible. The
appearance of a stable coexistent ferromagnetic-superconducting phase means, that
either Pauli limiting situation critical field has not been reached in the case of
spin-singlet pairing or else, the pairing has the spin-triplet nature, without
the component with spin $S^z=0$, and then the Pauli limit is not operative. 

The
present model with local spin-triplet pairing has its precedents of the same
type in the case of spin-singlet pairing, i.e., the Hubbard model with $U<0$
\cite{Micnas}, which played the central role in
singling out a nontrivial character of pairing in real space. Here, the same
role is being
played by the intra-atomic (but inter-orbital) ferromagnetic exchange. We
believe that this area of research unexplored so far in detail opens up new possibilities of studies of
weakly and moderately correlated magnetic superconductors \cite{Nomura2002}. 

The structure of this paper is as follows. In Section 2 we define the model and
the full Hartee-Fock-BCS approximation (i. e. mean field approximation for
magnetic
ordering with the concomitant BCS-type decoupling) for the coexistent
two-sublattice antiferromagnetic and spin-triplet superconducting phase (cf. also Appendix A for details). For completeness, in Appendix B we include also the analysis of a simpler coexistant superconducting-ferromagnetic phase. In
Section 3 we provide a detailed numerical analysis and construct the full phase
diagram on the Hund's rule coupling-band filling plane. We describe also the physical properties of the coexistent phases. In Appendix C we sketch a systematic approach of
going beyond Hartree-Fock approximation, i.e., including the spin fluctuations, starting from our Hartree-Fock-BCS state.

\section{Model and coexistent antiferromagnetic-spin-triplet
superconducting phase: mean-field-BCS approximation}\label{sec:model}
We start with the extended orbitally degenerate Hubbard Hamiltonian, which has the form
\begin{equation}
\begin{split}
\hat{H}&=\sum_{ij(i\neq j)ll^{\prime}\sigma}t^{ll^{\prime}}_{ij}
a_{il\sigma}^{\dag}a_{jl^{\prime}\sigma}+(U^{\prime}+J)\sum_{i}\hat{n}_{i1}\hat{n}_{i2}\\
&+U\sum_{il}\hat{n}_{il\uparrow}\hat{n}_{
il\downarrow}-J\sum_{ill^{\prime}(l\neq
l^{\prime})}\bigg(\mathbf{\hat{S}}_{il}\cdotp
\mathbf{\hat{S}}_{il^{\prime}}+\frac{3}{4}\hat{n}_{il}\hat{n}_{il^{\prime}}
\bigg),
\label{eq:H_start}
\end{split}
\end{equation}
where $l=1,2$ label the orbitals and the first term describes electron hopping
between atomic sites $i$ and $j$. For $l\neq l^{\prime}$ this term
represents electron hopping with change of the orbital (inter-site,
inter-orbital hybridization). Next two terms describe the Coulomb
interaction
between electrons on the same atomic site. However as one can see the second term contains the contribution that originates from the exchange interaction ($J$). The last term expresses the (Hund's
rule) ferromagnetic exchange between electrons localized on the same site, but
on different orbitals. This term is regarded as responsible for the local spin-triplet pairing in
the subsequent discussion. The components of the spin operator
$\mathbf{\hat{S}}_{il}=( \hat{S}^x_{il}, \hat{S}^y_{il}, \hat{S}^z_{il} )$ used
in (\ref{eq:H_start}) aquire the form
\begin{equation}
 \hat{S}^{x,y,z}_{il}=\frac{1}{2}\mathbf{\hat{h}}_{il}^{\dagger}\mathbf{
\sigma} _ {x,y,z}\mathbf{\hat{h}}_{il},
\end{equation}
where $\sigma_{x,y,z}$ are the Pauli matrices and
$\mathbf{h}_{il}^{\dagger}\equiv (a^{\dagger}_{il\uparrow},
a^{\dagger}_{il\downarrow})$. In our considerations we neglect the
interaction-induced intra-atomic singlet-pair hopping
($Ja^{\dag}_{i1\uparrow}a^{\dag}_{i1\downarrow}a_{i2\downarrow}a_{i2\uparrow}
+H.c.$) and the
correlation induced hopping ($Vn_{1\bar{\sigma}}(a^{\dagger}_{1\bar{\sigma}}a_{2\bar{\sigma}}+a^{\dagger}_{2\bar{\sigma}}a_{1\bar{\sigma}})+1\leftrightarrow2$ ) \cite{Nomura2002}, as they should not introduce any
important new qualitative feature in the
considered here spin-triplet paired states. What is more important, we assume that
$t^{12}_{ij}=t^{21}_{ij}$ and $t^{11}_{ij}=t^{22}_{ij}\equiv t_{ij}$, i.e., the
starting degenerate bands have the same width (\textit{the extreme degeneracy
limit}), as we are interested in establishing new qualitative features to the
overall phase diagram, that are introduced by the magnetic pairing. 

As has already been said, the aim of this work is to examine the spin-triplet
superconductivity coexisting with ferromagnetism and antiferromagnetism as
well
as the pure spin-triplet superconducting phase and the pure magnetically ordered
phases. Labels
defining the spin-triplet paired phases (A and A1) that are going to be used in
this work correspond to those defined for superfluid $^3$He according to the
Refs. \cite{Vollhardt} and \cite{Anderson}. Namely in the A phase the
superconducting gaps that correspond to Cooper pairs with total spin up and down
are equal ($\Delta_1=\Delta_{-1}\neq 0$, $\Delta_0=0$), whereas in the A1 phase
the only nonzero superconducting gap is the one that corresponds to the Cooper
pair with total spin up ($\Delta_1\neq 0$, $\Delta_{-1}=\Delta_0=0$). In this
section we show the method of calculations that is appropriate
for the superconducting phase coexisting with antiferromagnetism, as well as pure
superconducting phase of type A and pure antiferromagnetic phase. The
corresponding considerations for the case of ferromagnetically
ordered phases and superconducting phase A1 are deferred to the Appendix B.

From the start we make use of the fact that the full exchange
term can be represented by the real-space spin-triplet pairing operators, in the
following
manner 
\begin{equation}
J\sum_{ill^{\prime}(l\neq l^{\prime})}\bigg(\mathbf{\hat{S}}_{il}\cdotp
\mathbf{\hat{S}}_{il^{\prime}}+\frac{3}{4}\hat{n}_{il}\hat{n}_{il^{\prime}}
\bigg)\equiv 2J\sum_{i,m}\hat{A}^{\dagger}_{im}\hat{A}_{im},
\label{eq:Hund_pairing}
\end{equation}
which are of the form
\begin{equation}
\hat{A}^{\dagger}_{i,m}\equiv\left\{\begin{array}{cl}
a^{\dagger}_{i1\uparrow}a^{\dagger}_{i2\uparrow} & m=1,\\
a^{\dagger}_{i1\downarrow}a^{\dagger}_{i2\downarrow} & m=-1,\\
\frac{1}{\sqrt{2}}(a^{\dagger}_{i1\uparrow}a^{\dagger}_{i2\downarrow}+a^{\dagger}_{i1\downarrow}a^{\dagger}_{i2\uparrow}) & m=0.\\
\end{array}\right.
\label{eq: A_op}
\end{equation} 
Furthermore the inter-orbital Coulomb repulsion term can be expressed with the
use of spin-triplet pairing operators and the spin-singlet pairing operators in the following manner
\begin{equation}
 (U^{\prime}+J)\sum_{i}\hat{n}_{i1}\hat{n}_{i2}=(U^{\prime}+J)(\sum_{i}\hat{B}^{\dagger}
_i \hat{B}_i + \sum_{im}\hat{A}^{\dagger}_{im} \hat{A}_{im} ),
\label{eq:U_prime}
\end{equation}
where 
\begin{equation}
 \hat{B}^{\dagger}_i=\frac{1}{\sqrt{2}}(a^{\dagger}_{i1\uparrow}a^{\dagger}_{
i2\downarrow }
-a^{\dagger}_{i1\downarrow}a^{\dagger}_{i2\uparrow}),
\end{equation}
are the inter-orbital, intra-atomic spin-singlet pairing operators in real space.
Using Eq. (\ref{eq:Hund_pairing}) and (\ref{eq:U_prime}) one can write down
our model Hamiltonian in the form
\begin{equation}
\begin{split}
\hat{H}&=\sum_{ij(i\neq j)ll^{\prime}\sigma}t^{ll^{\prime}}_{ij}
a_{il\sigma}^{\dag}a_{jl^{\prime}\sigma}
+ U\sum_{il}\hat{n}_{il\uparrow}\hat{n}_{
il\downarrow}\\
&+(U^{\prime}+J)\sum_{i}\hat{B}_i^{\dagger}\hat{B}_i
-(J-U^{\prime})\sum_ {im}\hat{A}^{\dagger}_{im} \hat{A}_{im}.
\label{eq:H_start2}
\end{split}
\end{equation}
It should be noted here that for $J<U^{\prime}$, the inter-orbital Coulomb
repulsion suppresses the spin-triplet pairing mechanism and the superconducting
phases will not appear in the system in the weak-coupling (Hartree-Fock) limit.
For $3d$ electrons \cite{Sugano},
$U^{\prime}=U-2J$, thus the
necessary condition for the pairing to occur in our model
is $U<3J$. Usually, for 3d metals we have $U\sim 3J$, so it represents a rather stringent condition for the superconductivity to appear in that situation. We use this relation to fix the parameters for modeling purposes, not limited to 3d systems. This is also because e.g. 5f electrons in uranium compounds lead to a similar behavior as do 3d electrons. One should note that the considered here pairing is based on the intra-atomic inter-orbital ferromagnetic Hund's rule exchange. A simple extension to the situation with nonlocal J has been considered by X. Dai et al. \cite{Puetter2003}. Also, as we consider only weakly correlated regime, where the metallic state is stable, no orbital ordering is expected (cf. Klejnberg and Spa\l ek in \cite{Spalek2001}).

In our considerations the antiferromagnetic state represents the simplest
example of
the spin-density-wave state. In this state, we can divide our system into two
interpenetrating sublattices $A$ and $B$. The average staggered magnetic moment
of electrons on
each of the $N/2$ sublattice $A$ sites is equal, $<S_i^z>=<S^z_A> $, whereas
on the remaining $N/2$ sublattice $B$ sites we have $<S_i^z>=<S^z_B>\equiv
-<S^z_A>$.
In accordance with this division into two sublattices, we define different
annihilation operators for each sublattice, namely
\begin{equation}
a_{il\sigma}=\left\{\begin{array}{cl}
a_{il\sigma A} \quad \textrm{for i} \in A,\\
a_{il\sigma B} \quad \textrm{for i} \in B.\\
\end{array}\right.
\end{equation}
The same holds for the creation operators, $a^{\dagger}_{il\sigma}$. We assume that the charge ordering is absent. In this situation, we can write that
\begin{equation}
 <S^z_{i1A}>=<S^z_{i2A}>\equiv \bar{S}^z_{s}, \quad
 <S^z_{i1B}>=<S^z_{i2B}>\equiv -\bar{S}^z_{s},
\end{equation}
\begin{equation}
 <n_{i1A}>=<n_{i2A}>=<n_{i1B}>=<n_{i2B}>\equiv n/2, \quad
\end{equation}
where $n$ is the band filling.
In what follows, we treat the pairing and the Hubbard parts in the combined mean-field-BCS
approximation. In effect, we can write down the Hamiltonian transformed in reciprocal ($\mathbf{k}$) space in
the form:
\begin{widetext}
\begin{equation}
\begin{split}
\hat{H}_{HF}-\mu \hat{N}&=\sum_{\mathbf{k}l\sigma}\bigg[\epsilon_{\mathbf{k}}(a^
{ \dagger } _ { \mathbf{k}l\sigma
A}a_{\mathbf{k}l\sigma B} +
a^{\dagger}_{\mathbf{k}l\sigma B}a_{\mathbf{k}l\sigma
A})-\sigma I\bar{S}^z_s(\hat{n}_{\mathbf{k}l\sigma
A}-\hat{n}_{\mathbf{k}l\sigma B}) \bigg]\\
&+\sum_{kll^{\prime}(l\neq l^{\prime} \sigma)}
\epsilon_{\mathbf{k}12}(a^{\dagger}_{\mathbf{k}l\sigma
A}a_{\mathbf{k}l^{\prime}\sigma B} +
a^{\dagger}_{\mathbf{k}l\sigma B}a_{\mathbf{k}l^{\prime}\sigma
A})+\sum_{\mathbf{k}l\sigma}\bigg[\frac{n}{8}(U+2U^{\prime}
-J)-\mu\bigg ] (\hat { n } _ {
\mathbf{k}l\sigma A}+\hat{n}_{\mathbf{k}l\sigma B})\\
&+\sum_{\mathbf{k},m=\pm
1}(\Delta_{mA}^*\hat{A}_{\mathbf{k}mA}+\Delta_{mA}\hat{A}^{\dagger}_{\mathbf{k}mA} )+\sum_ {\mathbf{k},m=\pm
1}(\Delta_{mB}^*\hat{A}_{\mathbf{k}mB}+\Delta_{mB}\hat{A}^{\dagger}_
{ \mathbf{k}mB})\\
&+\sqrt{2}\sum_{\mathbf{k}}(\Delta_{0A}^*\hat{A}_{\mathbf{k}0A}+\Delta_{0A}\hat
{A}^{\dagger}_{\mathbf{k} 0A} )+\sqrt{2}\sum_
{\mathbf{k}}(\Delta_{0B}^*\hat{A}_{\mathbf{k}0B}+\Delta_{0B}\hat{A}^{\dagger}_
{ \mathbf{k}0B})-N\frac{n^2}{16}(U+2U^{\prime}
-J)\\
&+2NI(\bar{S}^z_s)^2-\frac{N}{2(J-U^{\prime})}(|\Delta_{1A}|^2+|\Delta_{-1A}
|^2+|\Delta_ { 1B }
|^2+|\Delta_{-1B}
|^2+2|\Delta_{0A}|^2+2|\Delta_{0B}|^2),
\end{split}
\label{eq:H_HF}
\end{equation}
\end{widetext}
where $I\equiv U+J$ is the effective magnetic coupling constant and
$\epsilon_{\mathbf{k}1}=\epsilon_{\mathbf{k}2}\equiv
\epsilon_{\mathbf{k}}$ is the dispersion relation. The results presented in the
next section have been carried out for square lattice with
nonzero hopping $t$ between nearest neighbors only. The corresponding bare
dispersion relation in a nonhybridized band acquires the form:
\begin{equation}
 \epsilon_{\mathbf{k}}=-2t\cos k_x - 2t\cos k_y.
\end{equation}
As we are considering the doubly degenerate band model situation, we make a
simplifying assumption that the hybridization matrix element
$\epsilon_{12\mathbf{k}}=\beta_h\epsilon_{\mathbf{k}}$, where $\beta_h \in
[0,1]$ is the parameter, which specifies the hybridization strength (i.e. represents a second scale of electron energies, in addition to $\epsilon_{\mathbf{k}}$). This means
that we have just one active atom per unit cell with a doubly degenerate orbital
of the same kind (their spatial asymmetry is disregarded). One should note
that the sums in (\ref{eq:H_HF}) (and in all
corresponding equations below) is taken over N/2 independent $\mathbf{k}$
states. In the Hamiltonian written above we have also introduced the
superconducting spin-triplet sublattice gap parameters 
\begin{equation}
\begin{split}
  \Delta_{\pm
1A(B)}&\equiv-\frac{2(J-U^{\prime})}{N}\sum_{\mathbf{k}}<\hat{A}_{\mathbf{k},
\pm 1A(B)}>,\\
\Delta_{0A(B)}&\equiv-\frac{2(J-U^{\prime})}{\sqrt{2}N}\sum_{\mathbf{k}}<\hat{A
} _{\mathbf {k},0A(B)}>.
\label{eq:pairing_def}
\end{split}
\end{equation}
The terms: $N\frac{n^2}{16}(U+2U^{\prime}-J)$ and
$\frac{n}{8}(U+2U^{\prime}-J)$ in (\ref{eq:H_HF} ) can be neglected, as
they lead only to a shift of
the reference point of the system energy. One should note that since
the real-space pairing mechanism is of intra-atomic nature, there is no direct
conflict with either ferro- or antiferro-magnetic ordering coexisting with it.   

\subsection{Antiferromagnetic (Slater) subbands}
The diagonalization of the Hamiltonian (\ref{eq:H_HF}) can be carried out in two
steps. In the first step we diagonalize the one particle part of the
Hartree-Fock Hamiltonian (the first two sums of (\ref{eq:H_HF})). Note that we have to carry out this step first, since we assume the bands are both hybridized and contain pairing part. By
introducing the four-composite fermion operator
$\mathbf{f}^{\dagger}_{\mathbf{k}\sigma}\equiv(a^{\dagger}_{
\mathbf{k}1\sigma
A},a^{\dagger}_{\mathbf{k}2\sigma A},
a^{\dagger}_{\mathbf{k}1\sigma
B},a^{\dagger}_{\mathbf{k}2\sigma B})$, we can express the one particle
Hamiltonian in the following form
\begin{equation}
\hat{H}^0_{HF}=\sum_{\mathbf{k}\sigma}
\mathbf{f}_{\mathbf{k}\sigma}^{\dagger}\mathbf{H}^0_{\mathbf{k}\sigma}
\mathbf{f}_{\mathbf{k}\sigma},
\end{equation}
where
$\mathbf{f}_{\mathbf{k}}\equiv(\mathbf{f}^{\dagger}_{\mathbf{k}})^{\dagger}$,
and
\begin{equation}
\mathbf{H}^0_{\mathbf{k}\sigma}=\left(\begin{array}{cccc}
-\sigma I\bar{S}^z_s & 0 & \epsilon_{\mathbf{k}} & \epsilon_{\mathbf{k}12}\\
0 & -\sigma I\bar{S}^z_s & \epsilon_{\mathbf{k}12} & \epsilon_{\mathbf{k}}\\
\epsilon_{\mathbf{k}} & \epsilon_{\mathbf{k}12} & \sigma I\bar{S}^z_s & 0\\
\epsilon_{\mathbf{k}12} & \epsilon_{\mathbf{k}} & 0 &\sigma I\bar{S}^z_s
\end{array} \right).
\label{eq:matrix_H}
\end{equation}
To diagonalize this Hamiltonian we introduce a generalized Bogoliubov transformation to new
operators $\tilde{a}_{kl\sigma A}$ and $\tilde{a}_{kl\sigma
B}$ in the following manner
\begin{equation}
\left(\begin{array}{cl}
        a_{\mathbf{k}1\sigma A}\\
        a_{\mathbf{k}2\sigma A}\\
        a_{\mathbf{k}1\sigma B}\\
        a_{\mathbf{k}2\sigma B}
       \end{array}\right)=\left(\begin{array}{cccc}
-U^+_{\mathbf{k}\sigma} & U^-_{\mathbf{k}\sigma} & V^+_{\mathbf{k}\sigma} &
-V^-_{\mathbf{k}\sigma}\\
-U^+_{\mathbf{k}\sigma} & -U^-_{\mathbf{k}\sigma} & V^+_{\mathbf{k}\sigma} &
V^-_{\mathbf{k}\sigma}\\
V^+_{\mathbf{k}\sigma} & -V^-_{\mathbf{k}\sigma} & U^+_{\mathbf{k}\sigma} &
-U^-_{\mathbf{k}\sigma}\\
V^+_{\mathbf{k}\sigma} & V^-_{\mathbf{k}\sigma} & U^+_{\mathbf{k}\sigma} &
U^-_{\mathbf{k}\sigma}\\
\end{array}\right)\left(\begin{array}{cl}
        \tilde{a}_{\mathbf{k}1\sigma A}\\
        \tilde{a}_{\mathbf{k}2\sigma A}\\
        \tilde{a}_{\mathbf{k}1\sigma B}\\
        \tilde{a}_{\mathbf{k}2\sigma B}
       \end{array}\right),
\label{eq:quasi1}
\end{equation}
where 
\begin{equation}
\begin{split}
U^{(\pm)}_{\mathbf{k}\sigma}=\frac{1}{\sqrt{2}}\bigg(1+\frac{\sigma
I\hat{S}^z_s}{\sqrt{ (\epsilon_ {\mathbf{k}}\pm
\epsilon_{\mathbf{k}12})^2+(I\bar{S}^z_s)^2}}\bigg)^{1/2},\\
V^{(\pm)}_{\mathbf{k}\sigma}=\frac{1}{\sqrt{2}}\bigg(1-\frac{\sigma
I\hat{S}^z_s}{\sqrt{
(\epsilon_ {\mathbf{k}}\pm
\epsilon_{\mathbf{k}12})^2+(I\bar{S}^z_s)^2}}\bigg)^{1/2}.
\label{eq:oznaczenia_U_V_A_S}
\end{split}
\end{equation}
One should note that the symbols $A$ and $B$ that appear as indexes of the
new quasi-particle operators $\tilde{a}_{\mathbf{k}l\sigma A}$
and $\tilde{a}_{\mathbf{k}l\sigma B}$, single out the new, hybridized,
quasi-particle sub-bands and do not correspond to the sublattices indices
$A$ and $B$
as in the case of operators $a_{\mathbf{k}l\sigma A}$
and $a_{\mathbf{k}l\sigma B}$. 
The dispersion relations in the new quasi-particle representation
acquire the form
\begin{equation}
 \begin{split}
  \tilde{\epsilon}_{\mathbf{k}1 A}&=-\sqrt{(\epsilon_
{\mathbf{k}}+\epsilon_{\mathbf{k}12})^2+(I\bar{S}^z_s)^2},\\
  \tilde{\epsilon}_{\mathbf{k}1 B}&=\sqrt{(\epsilon_
{\mathbf{k}}+\epsilon_{\mathbf{k}12})^2+(I\bar{S}^z_s)^2}, \\
  \tilde{\epsilon}_{\mathbf{k}2 A}&=-\sqrt{(\epsilon_
{\mathbf{k}}-\epsilon_{\mathbf{k}12})^2+(I\bar{S}^z_s)^2},\\
  \tilde{\epsilon}_{\mathbf{k}2 B}&=\sqrt{(\epsilon_
{\mathbf{k}}-\epsilon_{\mathbf{k}12})^2+(I\bar{S}^z_s)^2}.
 \end{split}
\end{equation}
As one can see, the new dispersion relations do not depend on the spin quantum numbers of the
quasi-particle. In general if
$\epsilon_{\mathbf{k}12}$ is not  $\sim \epsilon_{\mathbf{k}}$, we may have four
non-degenerate Slater subbands, which is not the case considered here. 
To express the pairing operators that are present in the Hamiltonian
(\ref{eq:H_HF}) in terms of the new quasi-particle operators, one
can use relations (\ref{eq:quasi1}) and the definitions (\ref{eq: A_op}).
The explicit form of the original pairing operators in terms
of the newly created quasi-particle operators is provided in 
Appendix A.

\subsection{Quasiparticle states for the coexistent antiferromagnetic and
superconducting phase}
In the second step of the diagonalization of (\ref{eq:H_HF}), a generalized
Nambu-Bogolubov-De Gennes scheme is
introduced to write down the complete H-F Hamiltonian again in the matrix
form, which allows for an easy determination of its eigenvalues. With the help
of
composite creation operator
$\mathbf{\tilde{f}}^{\dagger}_{\mathbf{k}\sigma}\equiv(\tilde{a}^{\dagger}
_{\mathbf{k}1\sigma A},\tilde{a}_{-\mathbf{k}
2\sigma A},\tilde{a}^{\dagger}_{\mathbf{k}1\sigma
B},\tilde{a}_{-\mathbf{k}2\sigma B})$, we can construct this new 4x4
Hamiltonian matrix and write
\begin{equation}
\hat{H}_{HF}-\mu\hat{N}=\sum_{\mathbf{k}\sigma}
\mathbf{\tilde{f}}_{\mathbf{k}\sigma}^{\dagger}\mathbf{H}_{\mathbf{k}\sigma}
\mathbf{\tilde{f}}_{\mathbf{k}\sigma}+2\sum_{\mathbf{k}}(\tilde{\epsilon}_{
\mathbf{k}2 A}+\tilde{\epsilon}_{\mathbf{k}2 B} ) -2\mu N+C,
\label{eq:H_HF_matrix}
\end{equation}
with
\begin{equation}
\mathbf{H}_{\mathbf{k}\sigma}\equiv\left(\begin{array}{cccc}
\tilde{\epsilon}_{\mathbf{k}1 A}-\mu & \delta_{1\mathbf{k}\sigma}
& 0 & \delta_{3\mathbf{k}\sigma}\\
\delta_{1\mathbf{k}\sigma}^* &
-\tilde{\epsilon}_{\mathbf{k}2 A}+\mu &
\delta_{4\mathbf{k}\sigma} & 0\\
0 & \delta_{4\mathbf{k}\sigma}^* & \tilde{\epsilon}_{\mathbf{k}1 B}-\mu &
\delta_{2\mathbf{k}\sigma}\\
\delta_{3\mathbf{k}\sigma}^* & 0 & \delta_{2\mathbf{k}\sigma}^* &
-\tilde{\epsilon}_{\mathbf{k}2 B}+\mu
\end{array} \right),
\label{eq:matrix_H2}
\end{equation}
and $\mathbf{\tilde{f}}_{\mathbf{k}}\equiv(\mathbf{\tilde{f}}^{\dagger}_{\mathbf{k}})^{\dagger}$.
The parameters $\delta_{l\mathbf{k}\sigma}$ are defined as follows
\begin{equation}
\begin{split}
 \delta_{1\mathbf{k}\sigma}&=\Delta_{\sigma
A}U^+_{\mathbf{k}\sigma}U^-_{\mathbf{k}\sigma}+\Delta_{\sigma
B}V^+_{\mathbf{k}\sigma}V^-_{\mathbf{k}\sigma},\\
 \delta_{2\mathbf{k}\sigma}&=\Delta_{\sigma
A}V^+_{\mathbf{k}\sigma}V^-_{\mathbf{k}\sigma}+\Delta_{\sigma
B}U^-_{\mathbf{k}\sigma}U^+_{\mathbf{k}\sigma},\\
 \delta_{3\mathbf{k}\sigma}&=-\Delta_{\sigma
A}U^+_{\mathbf{k}\sigma}V^-_{\mathbf{k}\sigma}+\Delta_{\sigma
B}U^-_{\mathbf{k}\sigma}V^+_{\mathbf{k}\sigma},\\
 \delta_{4\mathbf{k}\sigma}&=-\Delta_{\sigma
A}V^+_{\mathbf{k}\sigma}U^-_{\mathbf{k}\sigma}+\Delta_{\sigma
B}V^-_{\mathbf{k}\sigma}U^+_{\mathbf{k}\sigma}.
\end{split}
\end{equation}
Constant $C$ contains the last two terms of the r. h. s. of expression
(\ref{eq:H_HF}).
Hamiltonian (\ref{eq:H_HF_matrix}) and matrix (\ref{eq:matrix_H2}) have been
written under the assumption that $\Delta_{0
A}=\Delta_{0 B}\equiv 0$. Calculations for the more general case of nonzero gap
parameters for $m=0$ have been also done, but no stable coexisting
superconducting and
antiferromagnetic solutions have been found numerically. The only
coexisting solutions that have been found, fulfill the relation $\Delta_{0
A}=\Delta_{0 B}\equiv 0$. This fact can be understood by the following argument.
As in the antiferromagnetic state all lattice sites have nonzero magnetic
moment, the Cooper pairs in the spin-triplet state for $m=0$ (i.e. with the
total
spin
$S^z=0$, corresponding $<\hat{A}_{\mathbf{k}0}>$) are not
likely to appear. Nevertheless, we present the matrix form of the Hamiltonian
(\ref{eq:H_HF}) for the mentioned most general case, in Appendix A.
In our considerations here, we limit also to the situation with the real gap parameters
$\Delta_{\pm1A(B) }^*=\Delta_{\pm 1A(B)}$.
Then, the straightforward diagonalization of (\ref{eq:matrix_H2}) yields to the following Hamiltonian
\begin{equation}
\begin{split}
\hat{H}_{HF}-\mu\hat{N}&=\sum_{\mathbf{k}l\sigma}(-1)^{l+1}(\lambda_
{\mathbf{k}l\sigma A}
\alpha^{\dagger}_{\mathbf{k}l\sigma A}\alpha_{\mathbf{k}l\sigma A}\\
&+\lambda_
{\mathbf{k}l\sigma B}
\alpha^{\dagger}_{\mathbf{k}l\sigma B}\alpha_{\mathbf{k}l\sigma B})
+2\sum_{\mathbf{k}}(\tilde{\epsilon}_{\mathbf{k}2
A}+\tilde{\epsilon}_{\mathbf{k}2 B}
)\\
&+\sum_{\mathbf{k}\sigma}(\lambda_{\mathbf{k2\sigma
A}}+\lambda_{\mathbf{k2\sigma B}})-2\mu N+C,
\end{split}
\end{equation}
where $\lambda_{\mathbf{k}l\sigma A(B)}$ are the eigenvalues of the matrix
(\ref{eq:matrix_H2}) and $\alpha_{\mathbf{k}l\sigma
A(B)}$ ($\alpha^{\dagger}_{\mathbf{k}l\sigma A(B)}$) are the
quasi-particle annihilation (creation) operators, related to the
original annihilation and creation operators $\tilde{a}_{\mathbf{k}l\sigma}$,
$\tilde{a}^{\dagger}_{\mathbf{k}l\sigma}$ from the first step of our
diagonalization, via generalized Bogoliubov transformation of the form
\begin{equation}
\mathbf{\tilde{f}}_{\mathbf{k}\sigma}=\mathbf{U}^{\dagger}_{\mathbf{k}\sigma
}\mathbf{g}_{\mathbf{k}\sigma},
\end{equation}
with
$\mathbf{g}^{\dagger}_{\mathbf{k}\sigma}\equiv(\alpha^{\dagger}
_{\mathbf{k}1\sigma A},\alpha_{-\mathbf{k}
2\sigma A},\alpha^{\dagger}_{\mathbf{k}1\sigma
B},\alpha_{-\mathbf{k}2\sigma B})$. Eigenvectors of the
Hamiltonian matrix (\ref{eq:matrix_H2}) are the columns of the diagonalization
matrix $\mathbf{U}^{\dagger}_{\mathbf{k}}$. 
Using the definitions of gap parameters $\Delta_{\pm 1 A}$, $\Delta_{\pm 1 B}$,
the average number of particles per atomic site $n=\sum_l<\hat{n}_{il\uparrow
A}+\hat{n}_{il\downarrow A}>$, and the average magnetic moment per band per site
$\bar{S}^z=<\hat{n}_{il\uparrow A}-\hat{n}_{il\downarrow A}>/2$, we can
construct
the set of self-consistent equations for the mean-field parameters ($\Delta_{\pm
1A}$, $\Delta_{\pm 1B}$, $\bar{S}^z$) and for the chemical potential. The averages
that appear in the set of self-consistent equations,
$<\alpha^{\dagger}_{\mathbf{k}l\sigma A(B)}\alpha_{\mathbf{k}l\sigma A(B)}>$,
can
be replaced by the corresponding Fermi distribution functions
\begin{equation}
f((-1)^{l+1}\lambda_{\mathbf{k}l\sigma A(B)})=1/[\exp(\beta
(-1)^{l+1}\lambda_{\mathbf{k}l\sigma A(B)})+1],
\end{equation}
where $\beta=1/k_BT$. The eigenvalues and the
eigenvectors of (\ref{eq:matrix_H2}) are evaluated numerically while
executing the numerical procedure of
solving the set of self-consistent equations. For a given set of microscopic
parameters $n$, $J$, $U$, $U^{\prime}$ and temperature $T$, the set of
self-consistent
equations has several solutions that correspond to different phases. Free
energy can be evaluated for each of the solutions that have been found and the
one
that corresponds to the lowest value of the free energy is regarded as
the stable phase. The expression for the free energy functional in the
considered case has the form
\begin{equation}
\begin{split}
 F&=-\frac{1}{\beta}\sum_{\mathbf{k}l\sigma}\bigg[\ln \bigg(1+\exp
 (-\beta(-1)^{l+1}\lambda_{\mathbf{k}l\sigma A})\bigg)\\
&+\ln \bigg(1+\exp
 (-\beta(-1)^{l+1}\lambda_{\mathbf{k}l\sigma
B})\bigg)\bigg]\\
&+2\sum_{\mathbf{k}}(\tilde{\epsilon}_{\mathbf{k}2
 A}+\tilde{\epsilon}_{\mathbf{k}2 B}
 )+\sum_{\mathbf{k}\sigma}(\lambda_{\mathbf{k}2\sigma
 A}\\
&+\lambda_{\mathbf{k}2\sigma B})-\mu (2-n)N+C.
\end{split}
\end{equation}
Numerical results are carried out for square lattice with nonzero hopping $t$
between the nearest neighbors only.\newline
The described above numerical scheme is executed for the following selection
of phases:
\begin{itemize}
 \item normal state ({\bf NS}): $\Delta_{\pm1A(B)}=0$, $\bar{S}^z_s=0$
 \item pure superconducting phase type A ({\bf A}):
$\Delta_{\pm1A(B)}\equiv\Delta\neq0$, $\bar{S}^z_s=0$
 \item pure antiferromagnetic phase ({\bf AF}): $\Delta_{\pm1A(B)}=0$,
$\bar{S}^z_s\neq0$
 \item coexistent superconducting and antiferromagnetic phase ({\bf SC+AF}):
$\Delta_{\pm1A(B)}\neq0$, $\bar{S}^z_s\neq0$
\end{itemize}
The ferromagnetically ordered phases, that will also be included
in our
considerations in the following Sections, are listed below:
\begin{itemize}
 \item pure saturated ferromagnetic phase ({\bf SFM}): $\Delta_{\pm1A(B)}=0$,
$\bar{S}^z_u=\bar{S}^z_{u(max)}\neq0$
 \item pure nonsaturated ferromagnetic phase ({\bf FM}):
$\Delta_{\pm1A(B)}=0$, $0<\bar{S}^z_u<\bar{S}^z_{u(max)}$
 \item saturated ferromagnetic phase coexistent with superconductivity of type
A1 ({\bf A1+SFM}):\newline $\Delta_{1A(B)}\equiv \Delta_{1}\neq0$,
$\Delta_{-1A(B)}=0$, $\bar{S}^z_u=\bar{S}^z_{u(max)}\neq0$
 \item nonsaturated ferromagnetic phase coexistent with superconductivity of
type
A1 ({\bf A1+FM}):\newline $\Delta_{1A(B)}\equiv \Delta_{1}\neq0$,
$\Delta_{-1A(B)}=0$,
$0<\bar{S}^z_u<\bar{S}^z_{u(max)}$
\end{itemize}
It should be noted that $\bar{S}^z_u$ refers to the uniform magnetic moment per
band, per site in the ferromagnetically ordered phases, whereas
$\bar{S}^z_s$ is the
staggered magnetic moment that corresponds to the antiferromagnetic phases.
One could also consider the so called superconducting phase of type B for which
all superconducting gaps (including $\Delta_{0 A(B)}$) are equal and different
from zero. However this phase never coexists with magnetic ordering. What is
more important in the absence of magnetic ordering the superconducting phase A
has always lower free energy than the B phase. Therefore the superconducting B phase is absent
in the following discussion.


\section{Results and Discussion}\label{sec:reults}

We assume that $U^{\prime}=U-2J$ and $U=2.2J$, so
there are actually two independent parameters in the considered model - $n$ and
$J$.
The energies have been normalized to
the bare band-width $W=8|t|$, and $T$ expresses the reduced
temperature, $T\equiv k_BT/W$. 

\subsection{Overall phase diagram: coexistent magnetic-paired states}
\begin{figure}[h]
\centering
\begin{tabular}{cccc}
 (a) & & \quad \quad \quad  (b) & \\
  &\includegraphics[angle=0,width=0.3\textwidth]{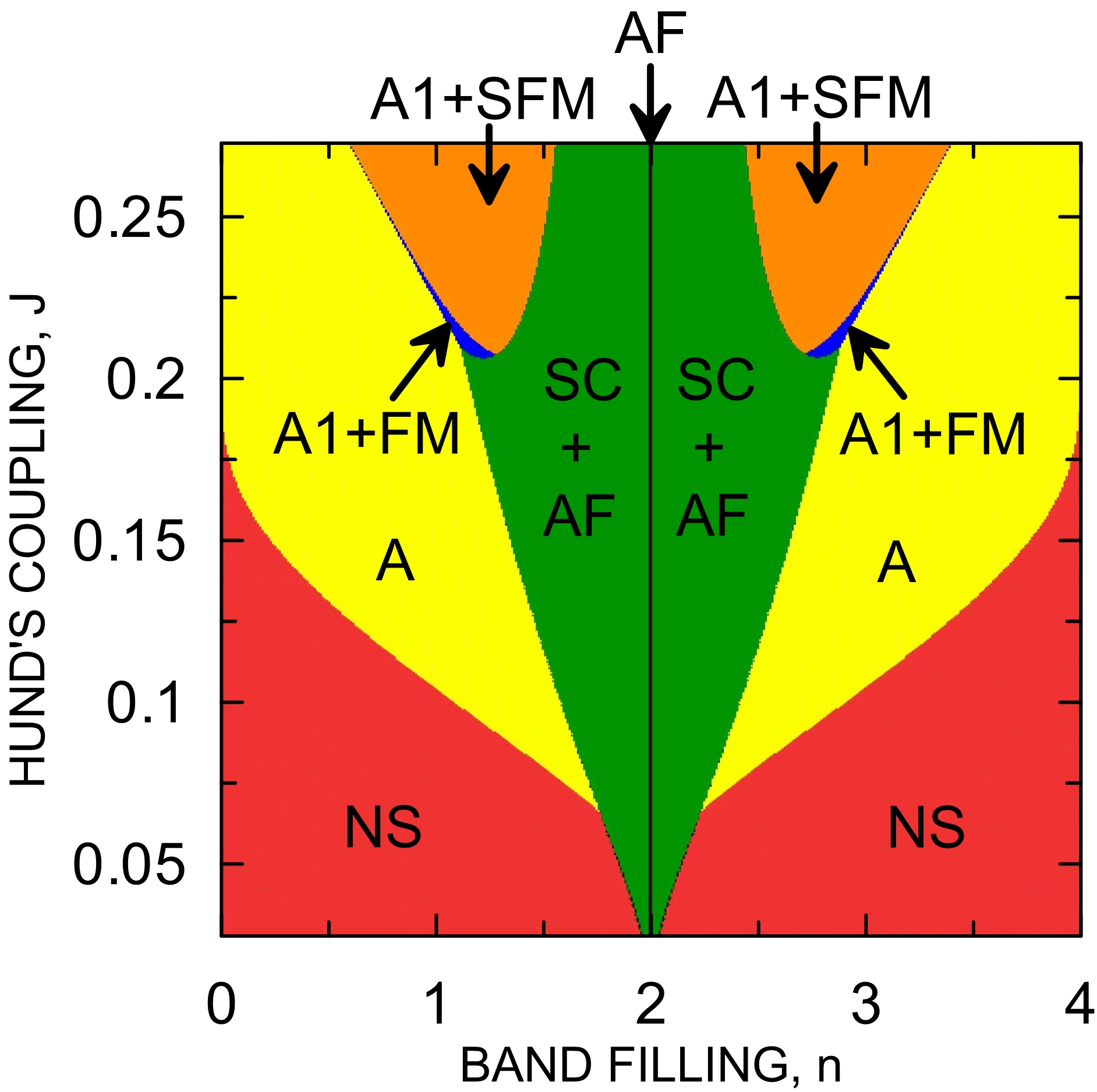} & &
\includegraphics[angle=0,width=0.3\textwidth]{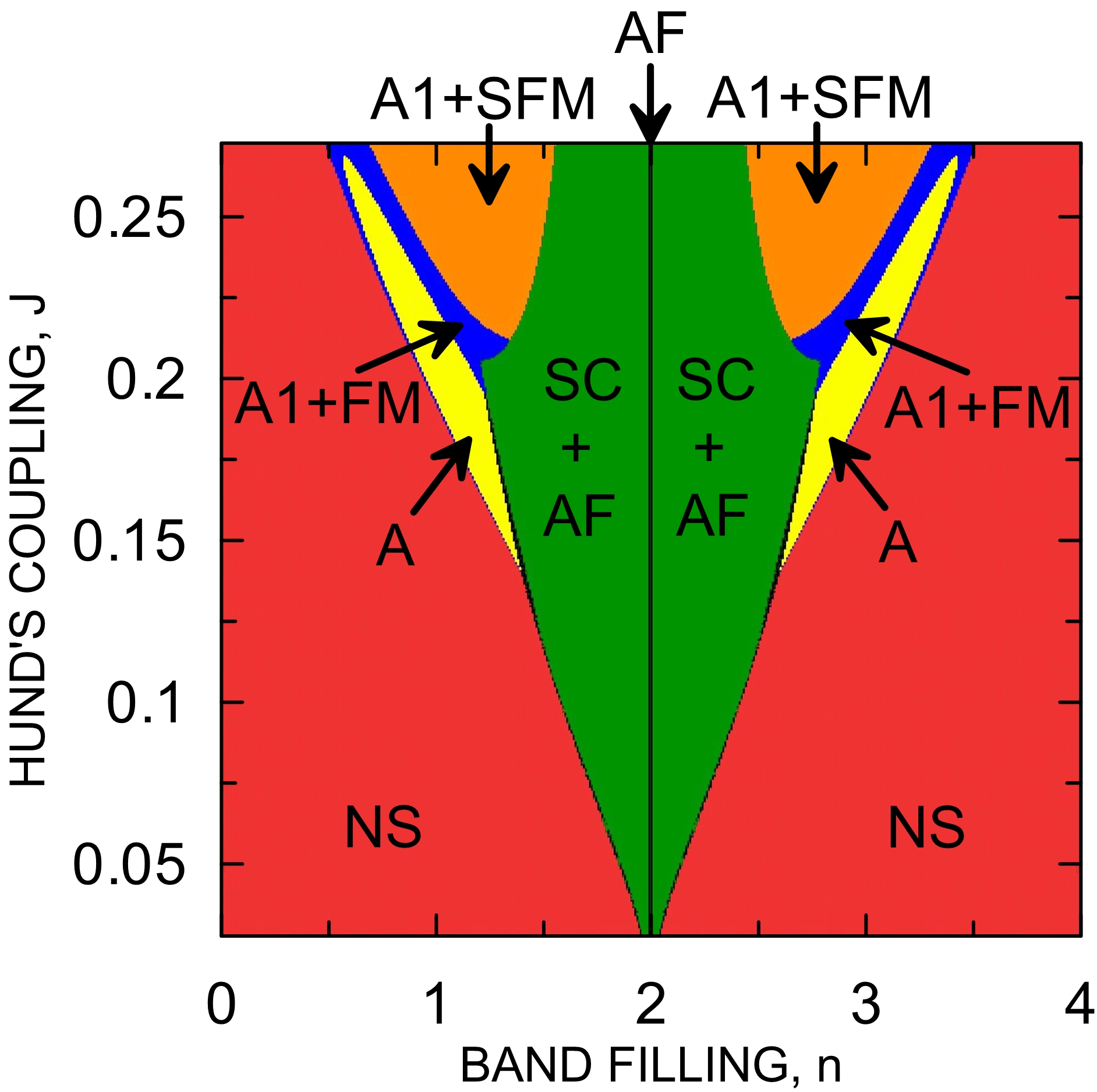}\quad \quad \quad \quad
 \\
 (c) & & \quad \quad \quad (d) & \\
  &\includegraphics[angle=0,width=0.3\textwidth]{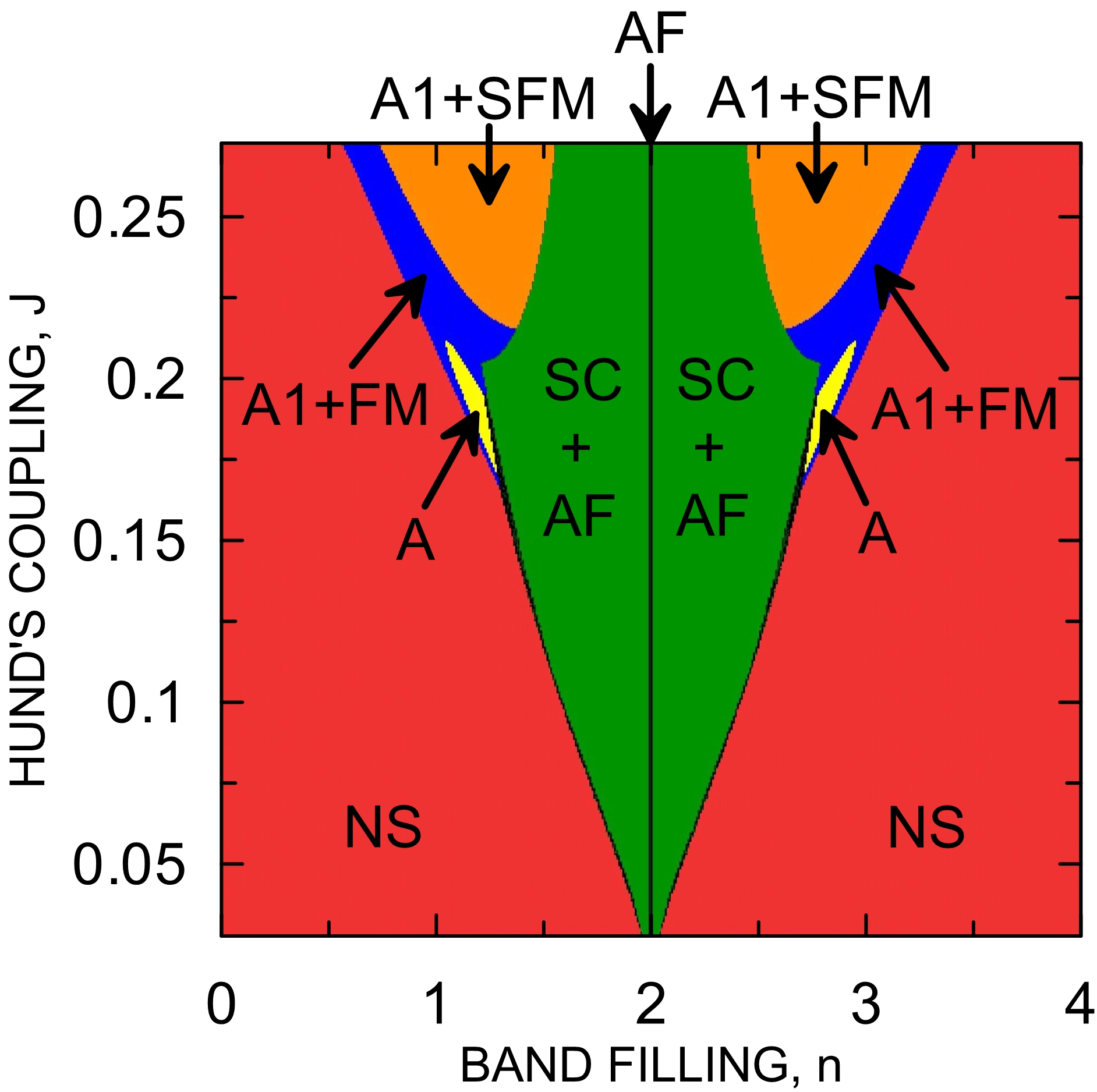} & &  
\includegraphics[angle=0,width=0.3\textwidth]{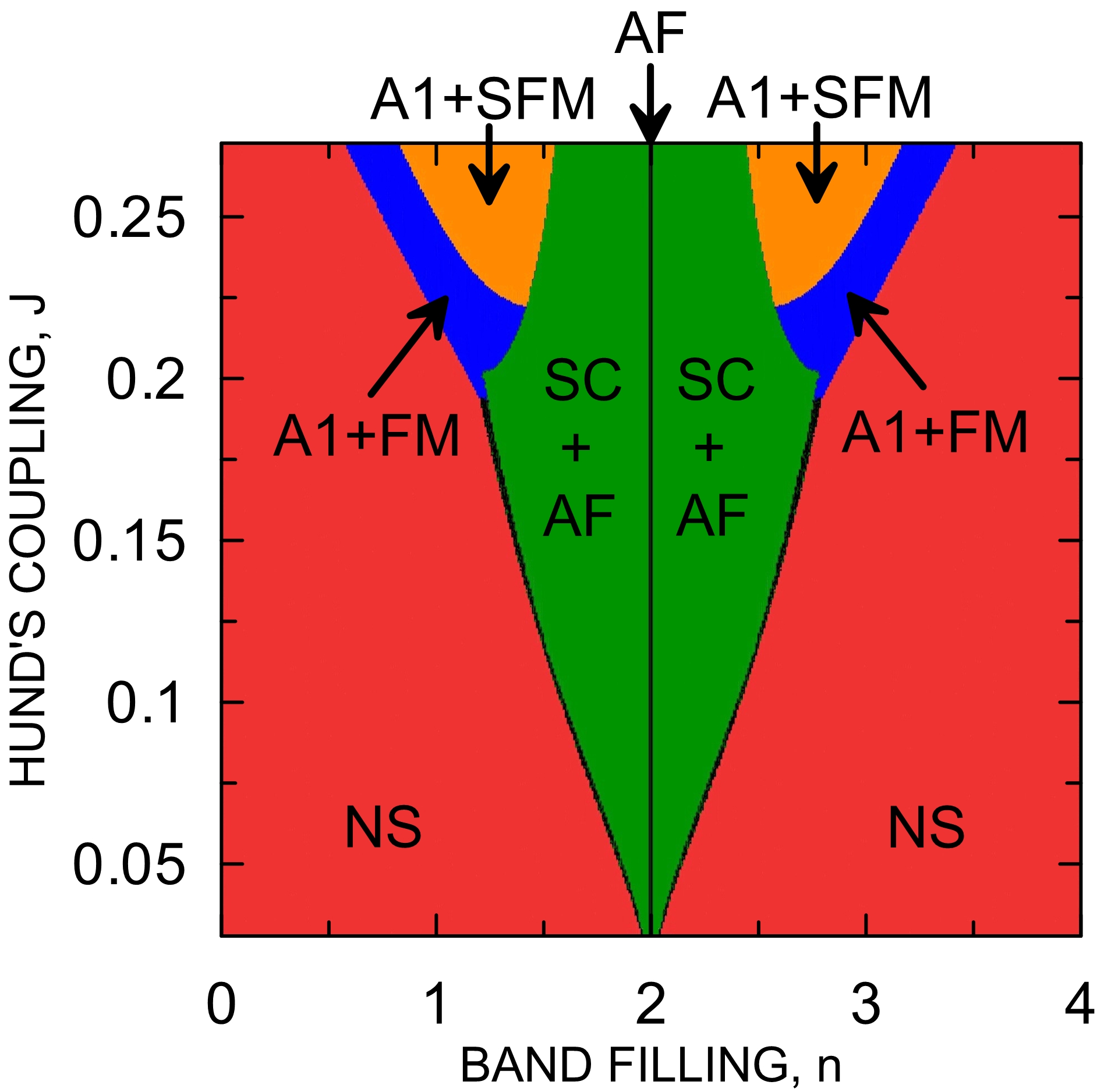}\quad \quad \quad
\quad 
\end{tabular}
\caption{(Color online) Phase diagrams in space $(n,J)$ for $T=10^{-4}$ and for
different values of the $\beta_h$
parameter: (a) $\beta_h=0.00$, (b) $\beta_h=0.04$, (c) $\beta_h=0.06$, (d)
$\beta_h=0.11$. Labels representing different phases are described in main
text. One sees that practically all magnetic phases here are in fact the
coexistent phases with superconductivity except the half filled situation where
we have pure AF phase}
\label{fig:diag_all}
\end{figure}

In Fig. \ref{fig:diag_all} a-d we present the complete phase
diagrams in coordinates $(n,J)$ for different values of the hybridization parameter $\beta_h$. They comprise sizeable regions of stable spin-triplet
superconducting phase coexisting with
either ferromagnetism or antiferromagnetism, as well as pure superconducting
phase A. In the phase SC+AF the calculated gap parameters fulfill the relations
\begin{gather}
\Delta_{+1A}=\Delta_{-1B}\equiv \Delta_{+},\nonumber\\
\Delta_{-1A}=\Delta_{+1B}\equiv \Delta_{-},\\
\Delta_{+}>\Delta_{-}.\nonumber
\end{gather}
For the singlet paired state one would have $\Delta_{+1A}=-\Delta_{-1A}$, which is not
the case here. For the case of half filled band, $n=2$, the superconducting gaps
$\Delta_+$
and $\Delta_-$ vanish and only pure (Slater type) AF survives. The appearance of the AF state for $n=2$ corresponds to the fact that the bare Fermi-surface topology has a rectangular structure with $Q=(\pi, \pi)$ nesting. This feature survives also for $\beta_h\neq 0$. Also, the symmetry of the phase diagrams with respect to half-filled band situation is a manifestation
of the
particle-hole symmetry, since the bare density of states is symmetric with
respect to the middle point of the band.
\onecolumngrid

\twocolumngrid
\noindent This feature of the problem provides an additional test for the correctness of the numerical results. It is clearly seen from the presented
figures that the influence of hybridization is significant quantitatively when it comes to the
superconducting phase A, as the region of its stability narrows down rapidly
with the increase of $\beta_h$.

\begin{figure}[htbp]
\centering
\begin{tabular}{ccc}
  (a) & & \quad \quad \quad \quad \quad\\
      & \includegraphics[angle=0,width=0.35\textwidth]{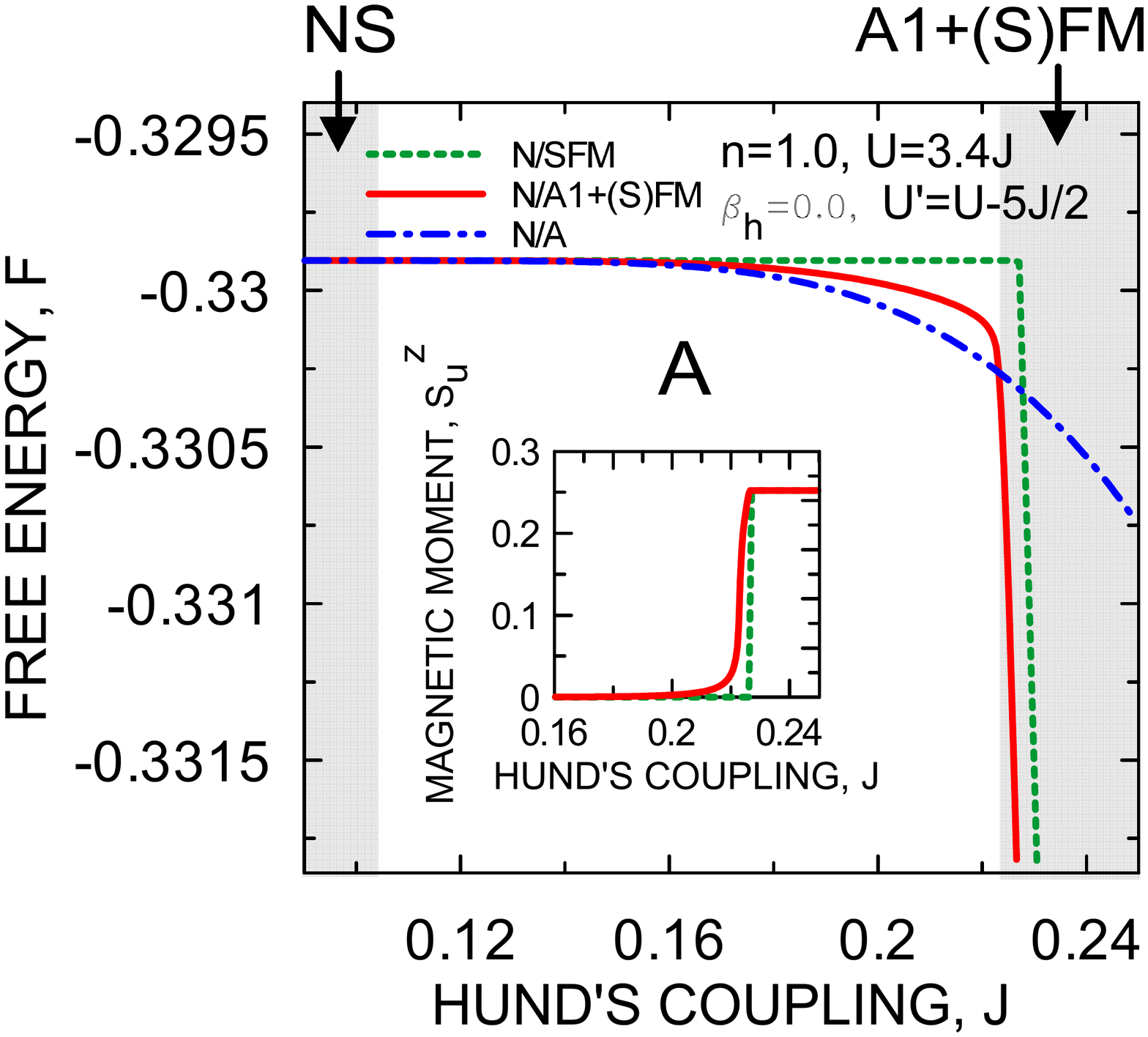} & \\
  (b) & & \quad \quad \quad \quad \quad\\
      & \includegraphics[angle=0,width=0.35\textwidth]{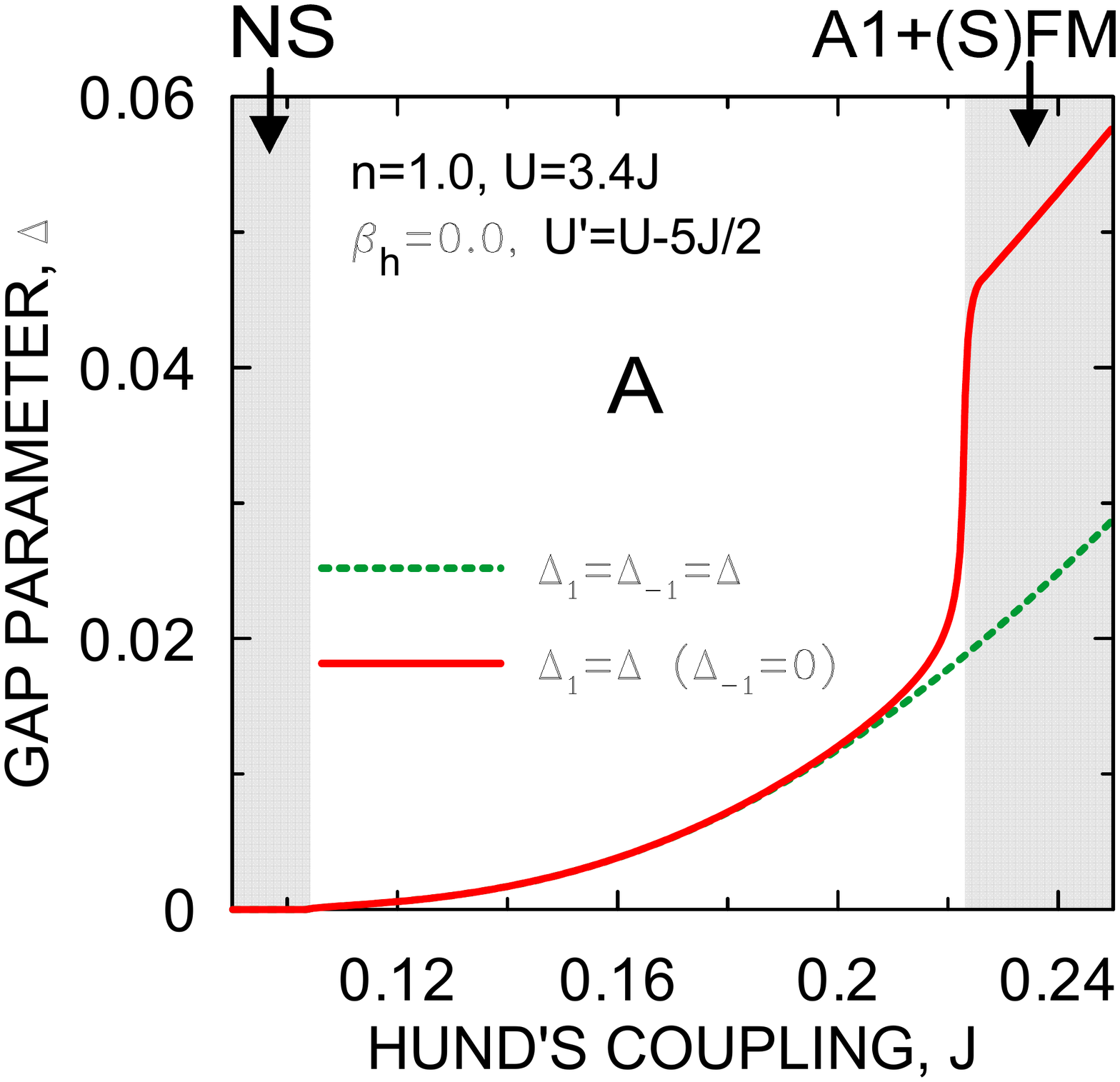} &\\
\end{tabular}
\caption{(Color online) Magnetic moment (per orbital per site), ground state energy and
superconducting gap as a function of $J$ near the Stoner
threshold for $n=1$ and $\beta_h=0.0$. Black vertical line in the inset marks the onset of saturated magnetism at the Stoner threshold.}
\label{fig:slice_J_b0}
\end{figure}

\begin{figure}[htpb]
\centering
\begin{tabular}{ccc}
  (a) & & \quad \quad \quad \quad \quad\\
      & \includegraphics[angle=0,width=0.35\textwidth]{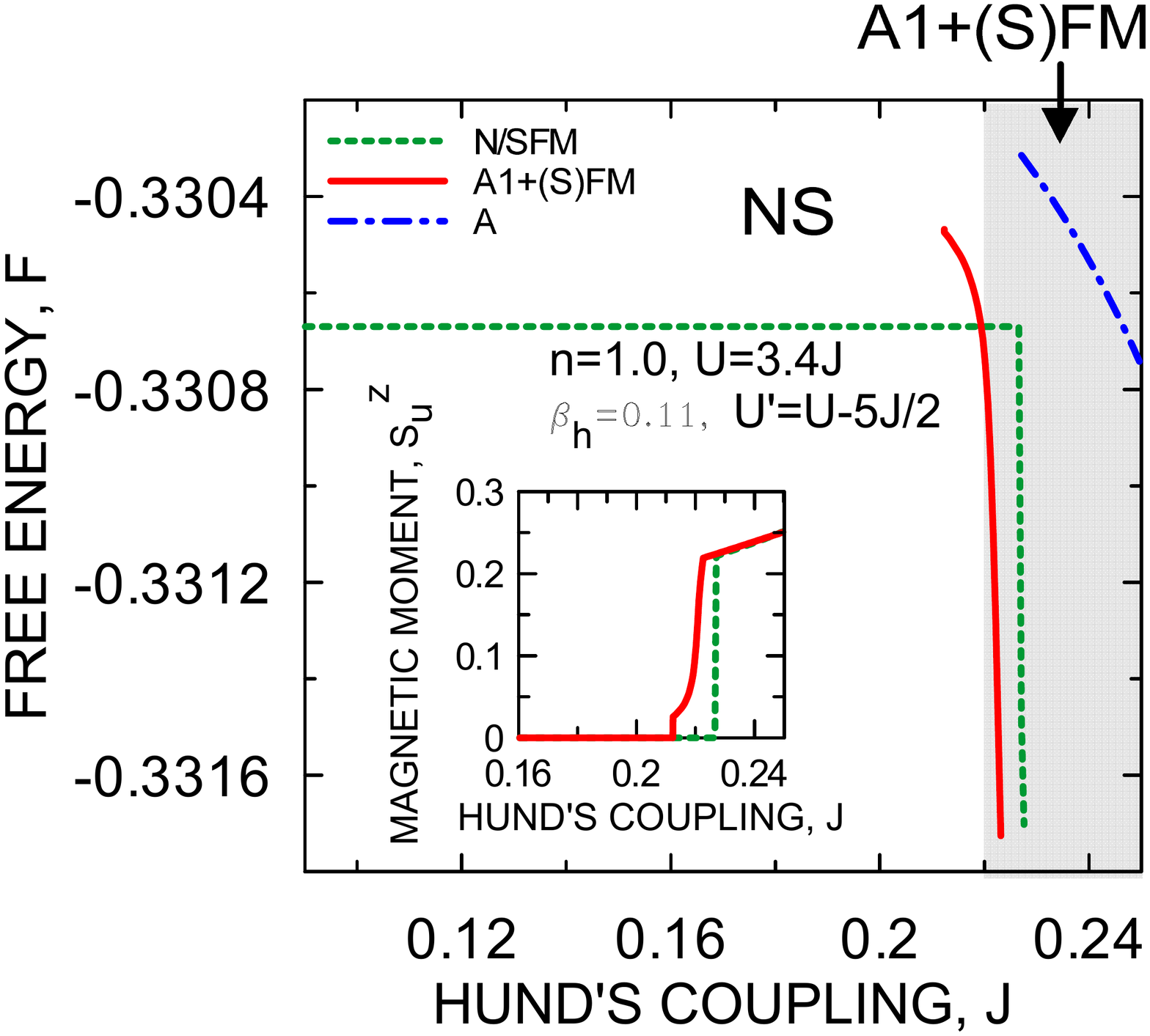} & \\
  (b) & & \quad \quad \quad \quad \quad\\
      & \includegraphics[angle=0,width=0.35\textwidth]{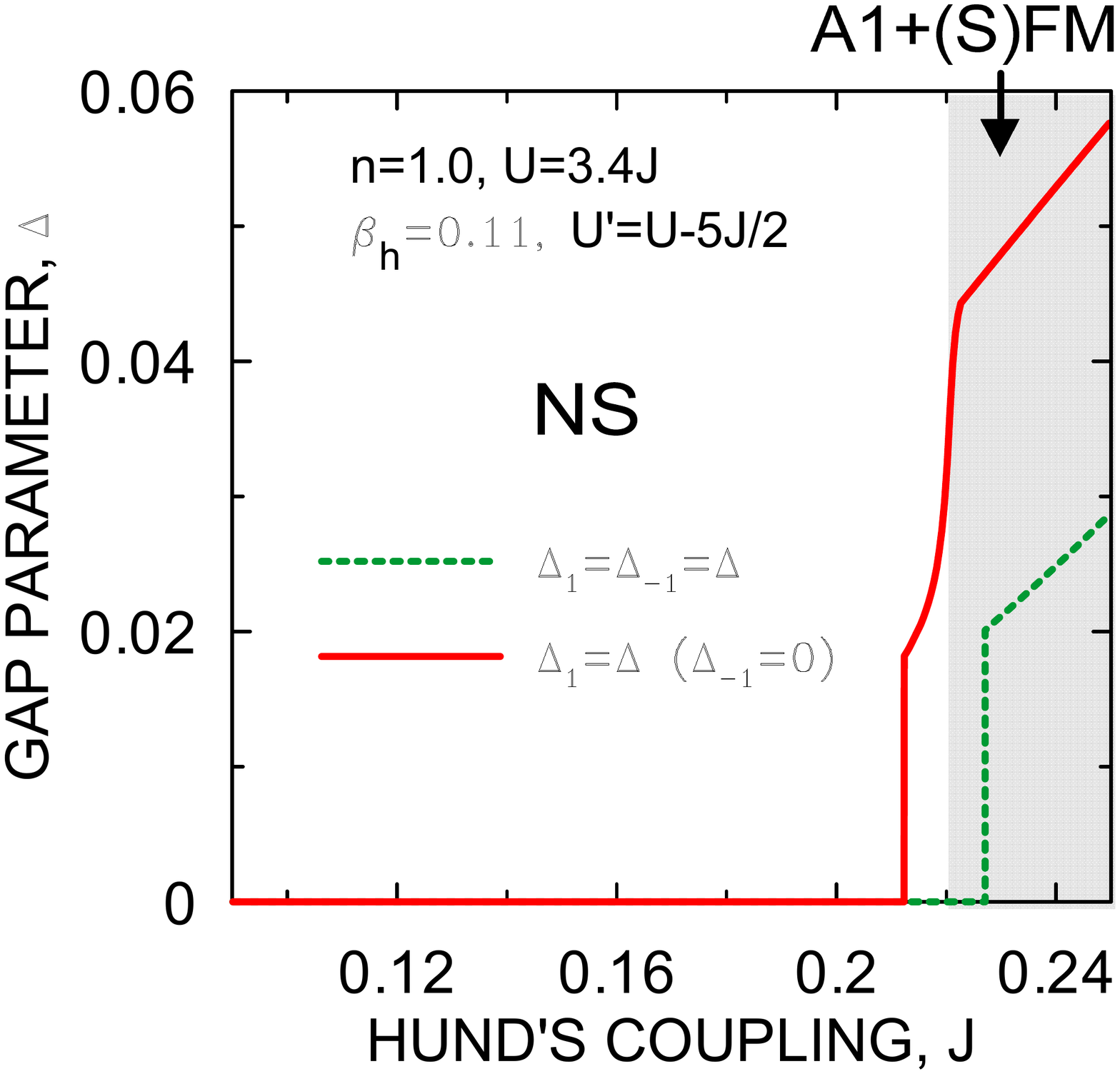} &\\
\end{tabular}
\caption{(Color online) Magnetic moment (per orbital per site), ground state energy and
superconducting gap as a function of $J$ near the Stoner
threshold for $n=1$ and $\beta_h=0.11$. Black vertical line in the inset marks the onset of  magnetism at the Stoner threshold.}
\label{fig:slice_J_b011}
\end{figure}

The stability areas of A1+FM and NS phases
expand on the expense of A and A1+SFM phases. With the further
increase of the hybridization, the
stability of A phase is completely suppressed, as shown in Fig.
\ref{fig:diag_all} d. The regions
of stable antiferromagnetically ordered phase do
not alter significantly with the increasing hybridization. 
\begin{figure}[htbp]
\centering
\begin{tabular}{ccc}
  (a) & & \quad \quad \quad \quad \quad\\
      & \includegraphics[angle=0,width=0.3\textwidth]{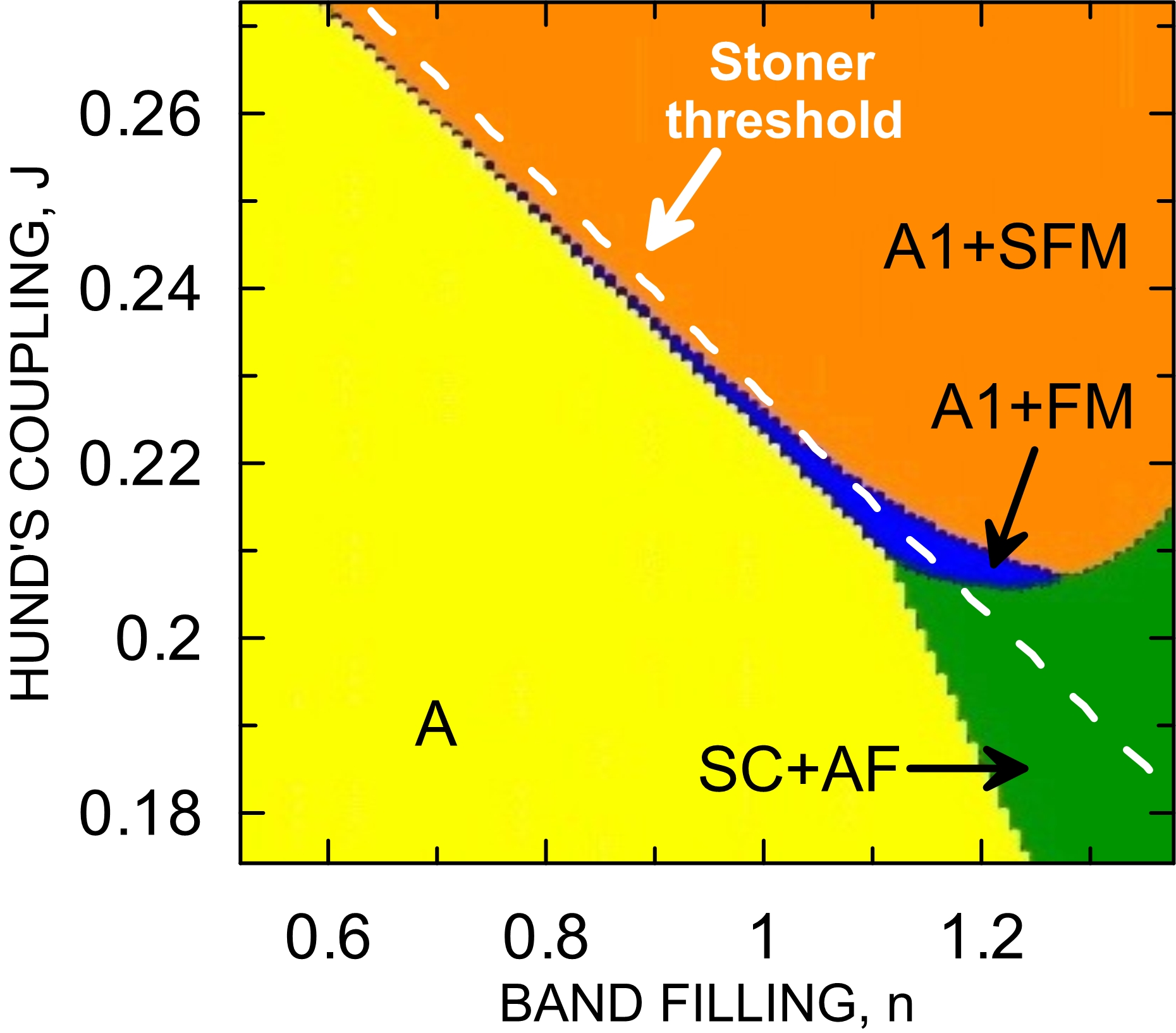} & \\
  (b) & & \quad \quad \quad \quad \quad\\
      & \includegraphics[angle=0,width=0.3\textwidth]{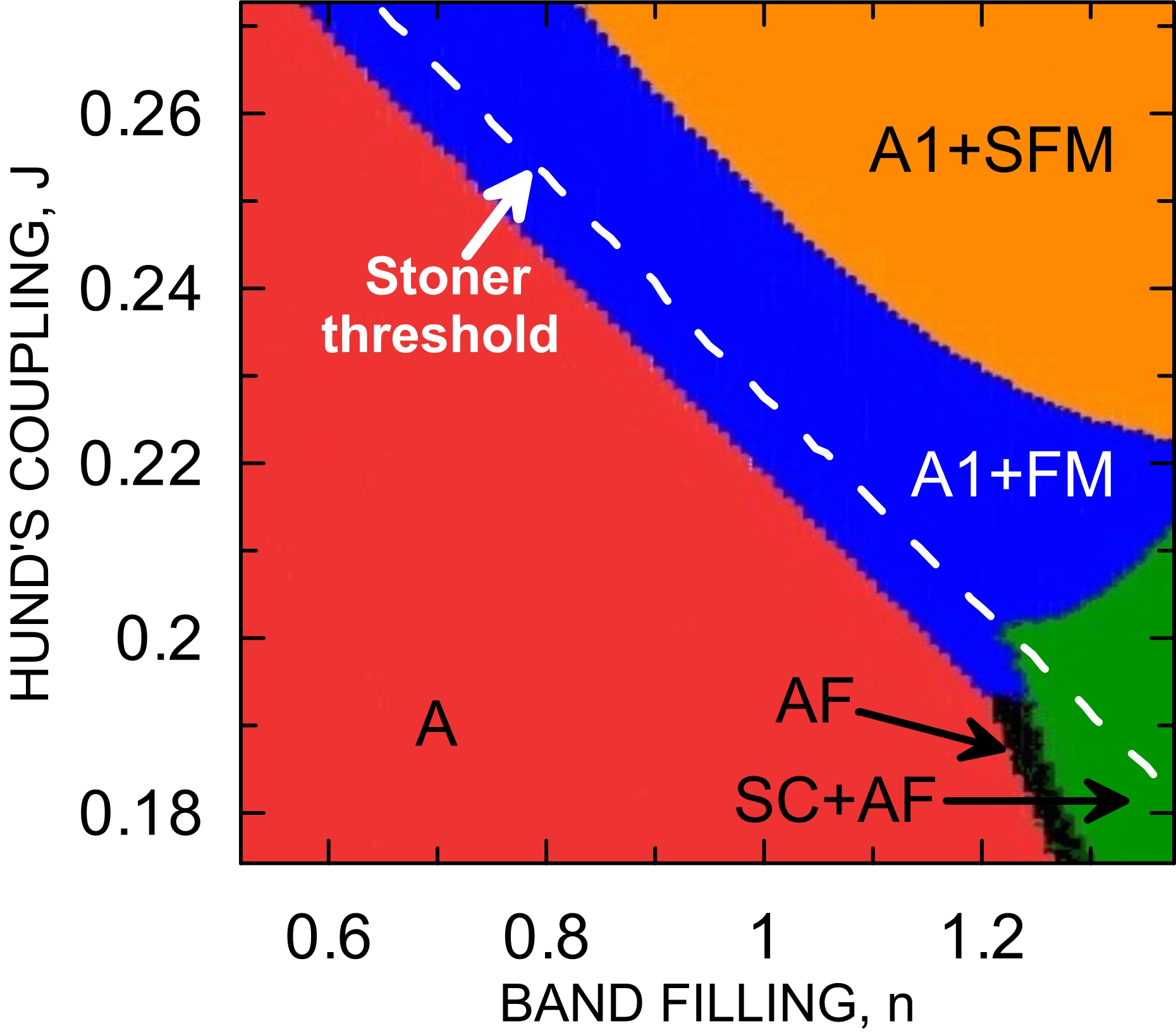} &\\
\end{tabular}
\caption{(Color online) Phase diagrams zoomed in space $(n,J)$ with the dashed
line marking the Stoner threshold for the onset of pure ferromagnetism. The values of the hybridization parameter are $\beta_h$: (a)
$\beta_h=0.00$, (b) $\beta_h=0.11$, while the temperature is $T=10^{-4}$.}
\label{fig:diag_stoner}
\end{figure}
To relate the appearance of superconductivity with the onset of ferromagnetism we
have marked explicitly in Fig. \ref{fig:diag_stoner} the Stoner threshold on the phase diagram. One sees clearly that only the A1 phase appearance
is related to the onset of ferromagnetism. What is more important, the FM phase coexisting
with the paired A1 phase, becomes stable for slightly lower $J$ values than the Stoner threshold for appearance of pure FM phase. The A1+FM coexistence near the Stoner threshold can be analyzed by showing
explicitly the magnetization and superconducting gap evolution with increasing
$J$. This is shown in Figs. \ref{fig:slice_J_b0} and \ref{fig:slice_J_b011}. One sees explicitly that the
nonzero magnetization appears slightly below the Stoner threshold and is thus
induced by the onset of A1 paired state. In other words, superconductivity
enhances magnetism. But opposite is also true, i.e., the gap increases rapidly
in this regime, where magnetization changes. The situation is preserved for
nonzero hybridization. The transition A$\rightarrow$A1+FM is sharp, as
detailed free-energy plot shows. For $\beta_h=0.11$ in a certain range of $J$
the superconducting solutions A1+FM and A cannot be found by the numerical
procedure. That is why the curves representing the gap parameters $\Delta$ and
free energy suddenly break. The most important and surprising conclusion is
that in A1+FM phase only the electrons in spin-majority subband are paired. This
conclusion may have important practical consequences for spin filtering across
NS/A1+FM interface, as discussed at the end. Nevertheless, one should note that
the partially polarized (FM) state appears only in a narrow window of $J$ values
near the Stoner threshold, at least for the selected density of states.


Summarizing, we have supplemented the well known magnetic phase diagrams with
the appropriate stable and spin-triplet paired states. A relatively
weak hybridization of band states destabilizes pure paired states but
stabilizes coexistent superconducting-magnetic phases except for the half-filled
band case, when the appearance of the Slater gap at the Fermi level excludes any
superconducting state. A very interesting phenomenon of pairing for one-spin
(majority) electrons occurs near the Stoner threshold for the onset of FM phase
and extends to the regime slightly below threshold.

\subsection{Detailed physical properties}

\begin{figure}[htpb]
\centering
\begin{tabular}{ccc}
  (a) & & \quad \quad \quad \quad \quad\\
      & \includegraphics[angle=-90,width=0.35\textwidth]{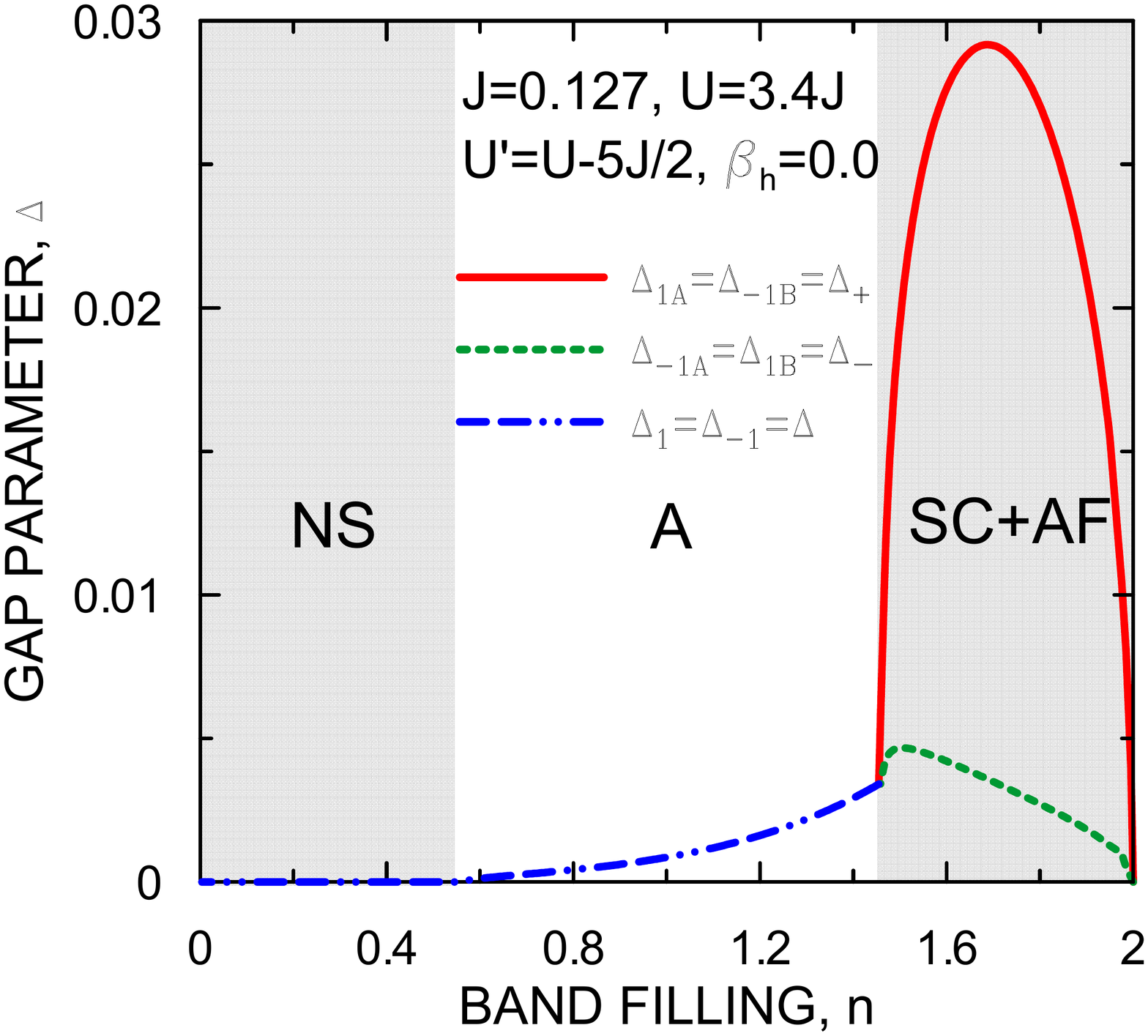} & \\
  (b) & & \quad \quad \quad \quad \quad\\
      & \includegraphics[angle=-90,width=0.35\textwidth]{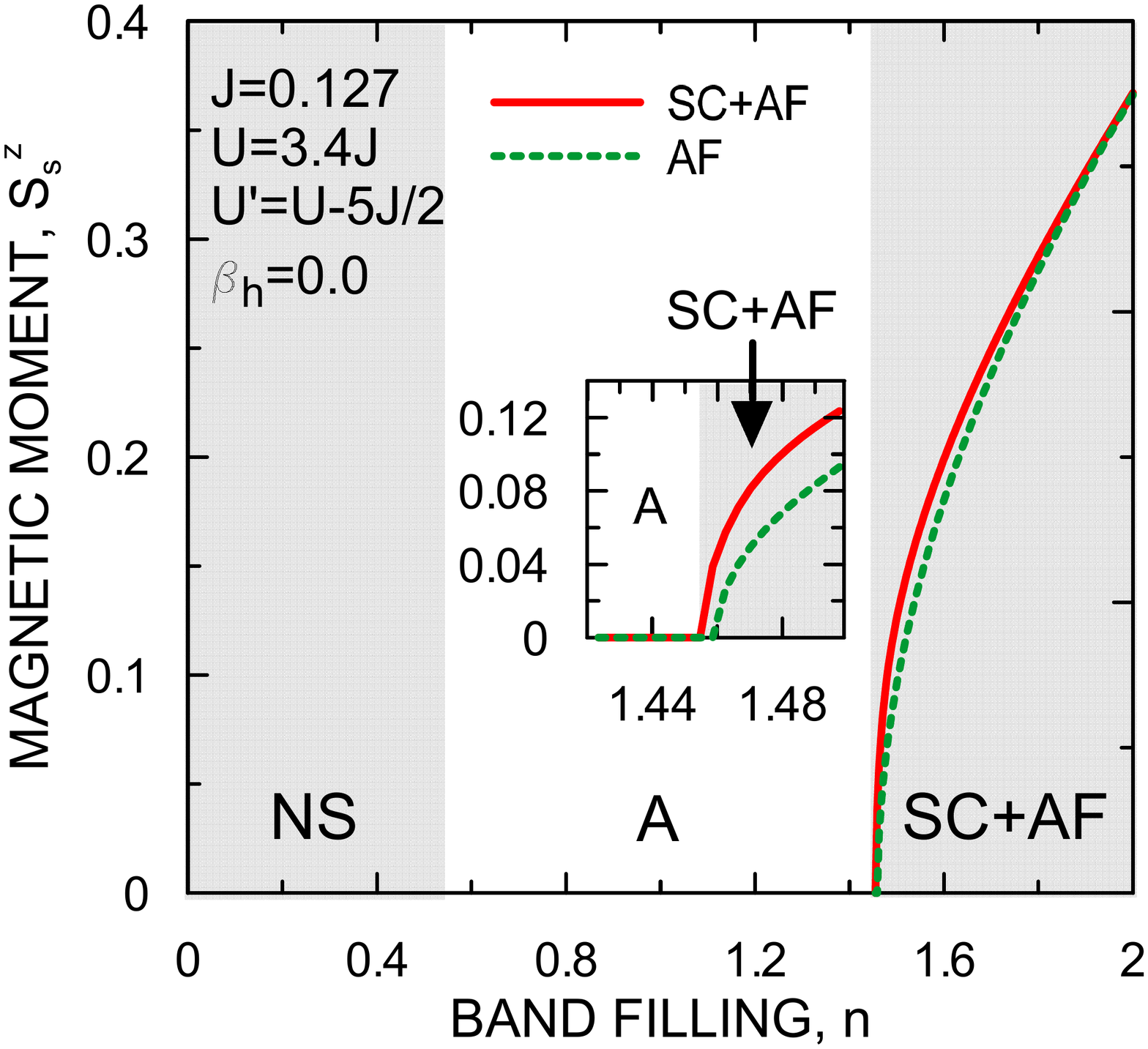} &\\
\end{tabular}
\caption{(Color online) Low temperature values of the superconducting gaps
and the staggered
magnetic moment both as a function of band filling for $\beta_h=0$ and
$J=0.175$. The stable
phases are appropriately labelled in the regimes of their stability. Note that
$\Delta_-<<\Delta_+$, i.e., the paired state is closer to A1 state than to A
state in the coexistent regime.}
\label{fig:slice_n_b0}
\end{figure}

\begin{figure}[htpb]
\centering
\begin{tabular}{ccc}
  (a) & & \quad \quad \quad \quad \quad\\
      & \includegraphics[angle=-90,width=0.37\textwidth]{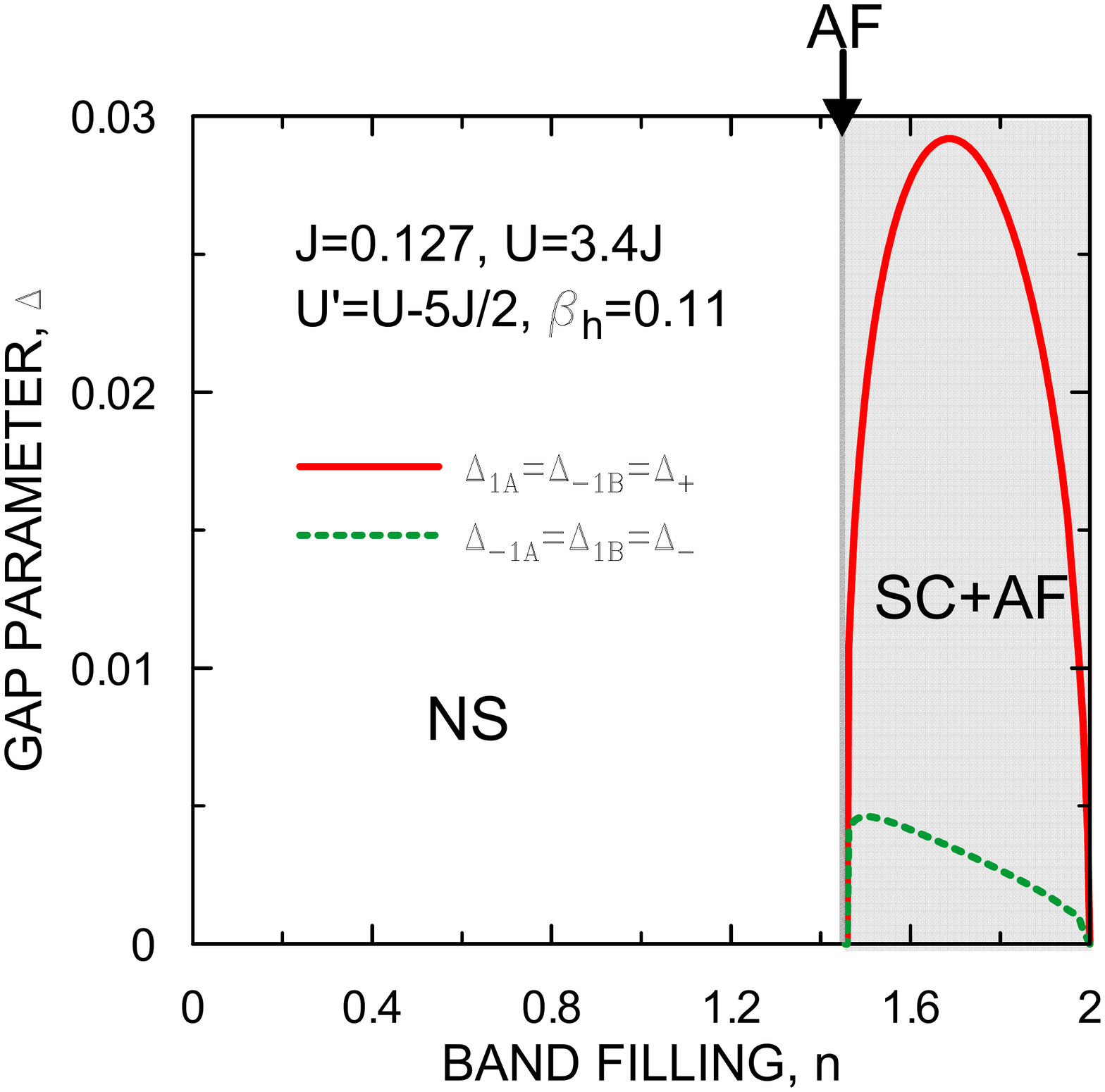} & \\
  (b) & & \quad \quad \quad \quad \quad\\
      & \includegraphics[angle=-90,width=0.37\textwidth]{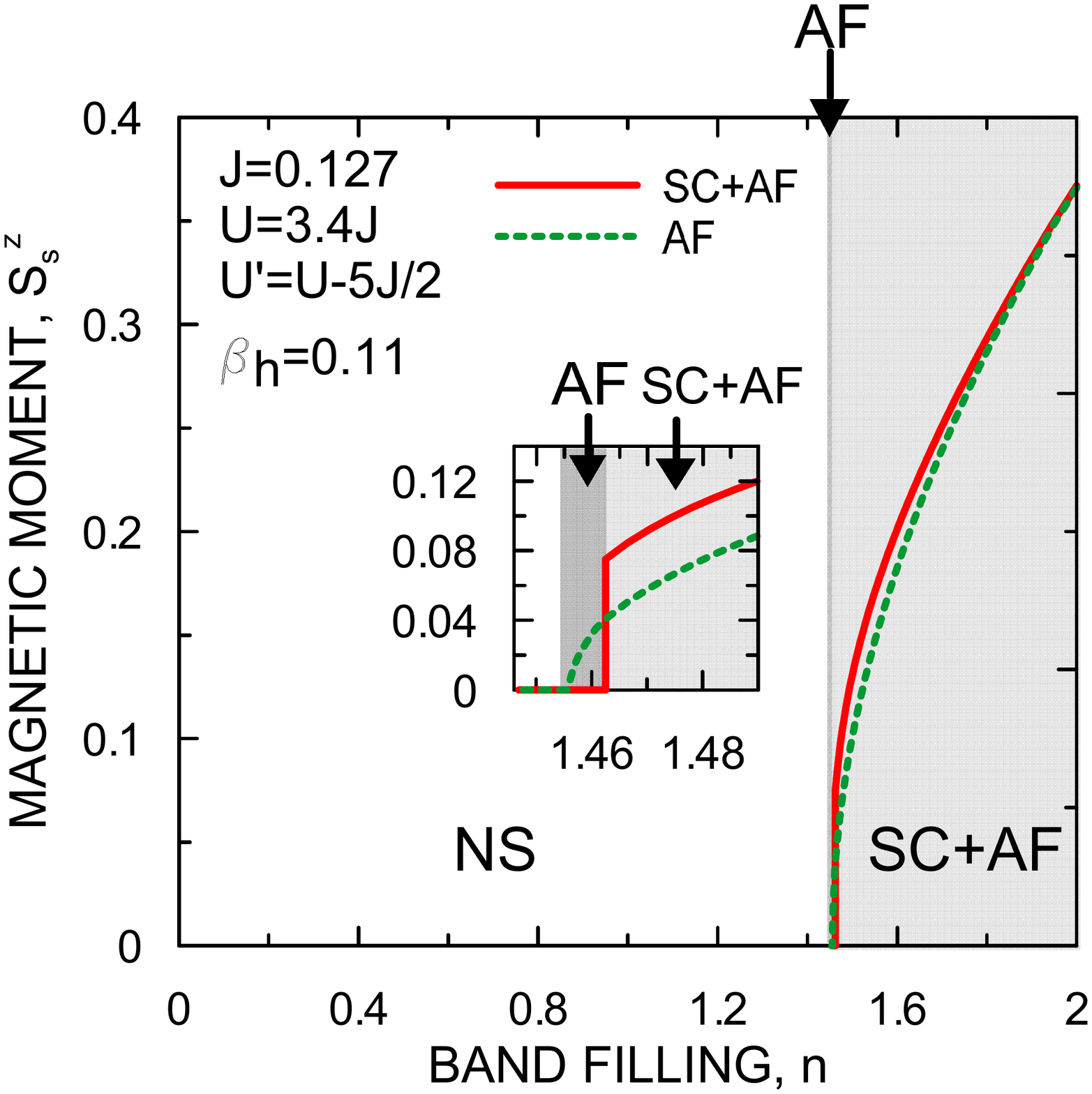} &\\
\end{tabular}
\caption{(Color online) Low temperature values of the superconducting gaps
and the staggered
magnetic moment as a function of band filling for $\beta_h=0.11$ and $J=0.175$.
Note the
disappearance of the pure A phase and that again $\Delta_-<<\Delta_+$. The
inset in (b) illustrates the fact that, a pure AF phase appears in a
very narrow regime of $n$ before the SC+AF phase becomes stable, whereas the inset in (a) shows the free energy of those two phases for $n$ close to $n_C$ when a weak first-order transition occurs.}
\label{fig:slice_n_b11}
\end{figure}

In Figs. \ref{fig:slice_n_b0} and
\ref{fig:slice_n_b11} we show the low-temperature values of superconducting gaps
and the staggered magnetic moment as a function of band filling. In the SC+AF
phase both gap parameters $\Delta_+$ and $\Delta_-$ decrease continuously to
zero as
the system approaches the half filling. On the contrary, the staggered
magnetic
moment $\bar{S}^z_s$ reaches then the maximum. For the case of $\beta_h=0.0$,
below
the critical value of band filling, $n_c\approx 1.45$ , the gap parameters
$\Delta_{+}$ and $\Delta_{-}$ are equal and the staggered magnetic moment
vanishes. In this regime the superconducting phase of type A is the stable one.
For A phase the superconducting gap decreases with the band-filling
decrease and becomes zero for some particular value of $n$. Below that
value the NS (paramagnetic state) is stable. It is clearly seen that the
appearance of two gap parameters above
$n_c$ is
connected with the onset of the staggered-moment structure, as above $n_c$
we have $S_s^z\neq 0$ (cf. Fig. \ref{fig:slice_n_b0}b). For comparison, we also
show the
staggered moment for pure AF in Figs. \ref{fig:slice_n_b0}b and
\ref{fig:slice_n_b11}b (dashed line). As one can see, the appearance of SC
increases
slightly the
staggered moment in SC+AF phase.  
For $\beta_h=0.11$ below some critical value of band filling $n_c\approx 1.473$
in a very narrow range of $n$, a pure AF phase is stable. The inset in Fig.
\ref{fig:slice_n_b11}a shows that there is aweak first-order transition between the
AF+SC and SC phases as a function of doping. The A phase is not stable in this
case.

One should mention that the easiness, with which the superconducting triplet
state is accommodated within the antiferromagnetic phase stems from the fact
that the SC gaps have an intra-atomic origin and the corresponding spins have
then the tendency to be parallel. Therefore, the pairs
respect the Hund's rule and do not disturb largely the
staggered-moment structure, which is of interatomic character. 

\begin{figure}[htpb]
\centering
\begin{tabular}{ccc}
  (a) & & \quad \quad \quad \quad \quad\\
      & \includegraphics[angle=-90,width=0.35\textwidth]{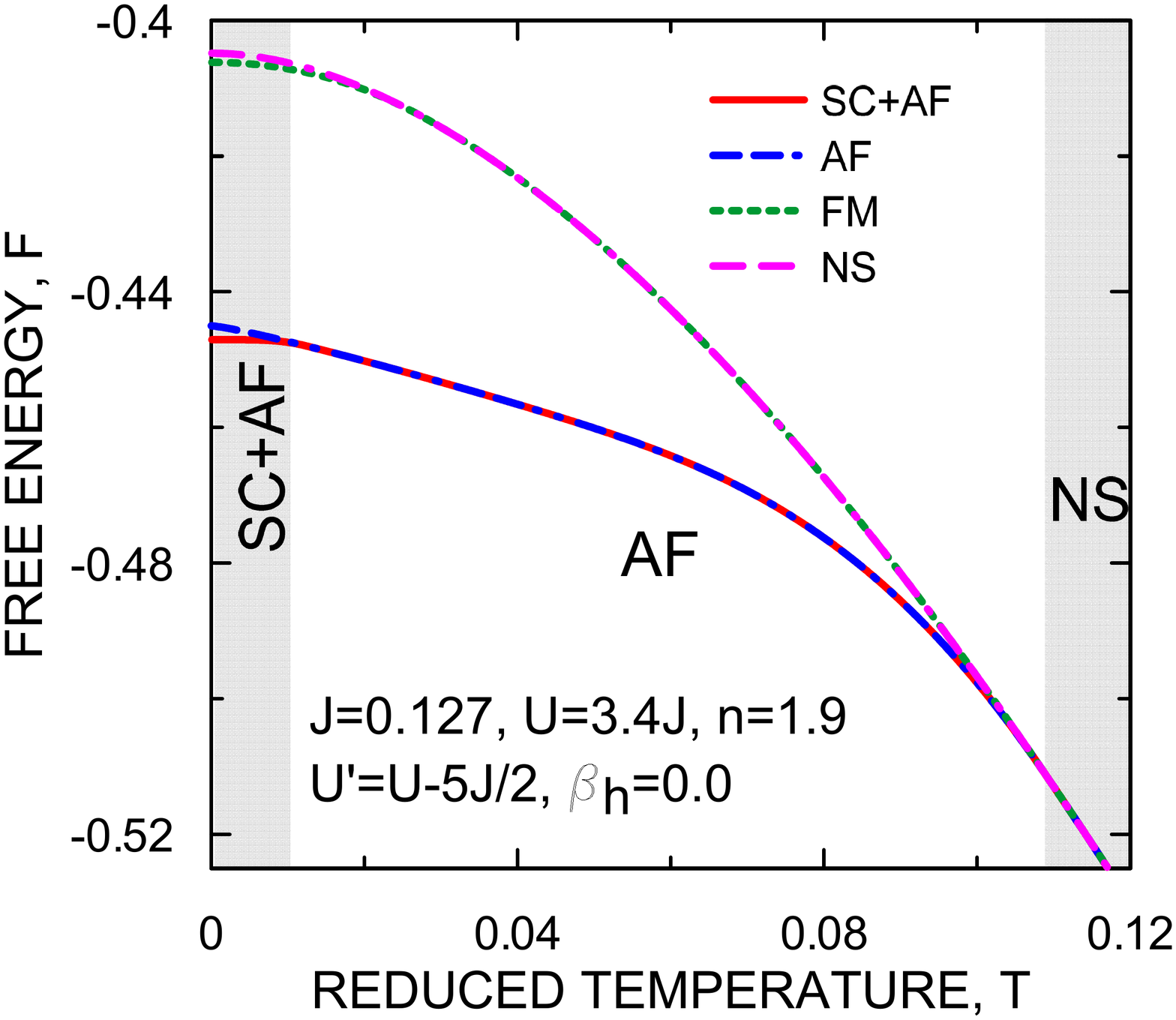} & \\
  (b) & & \quad \quad \quad \quad \quad\\
      & \includegraphics[angle=-90,width=0.35\textwidth]{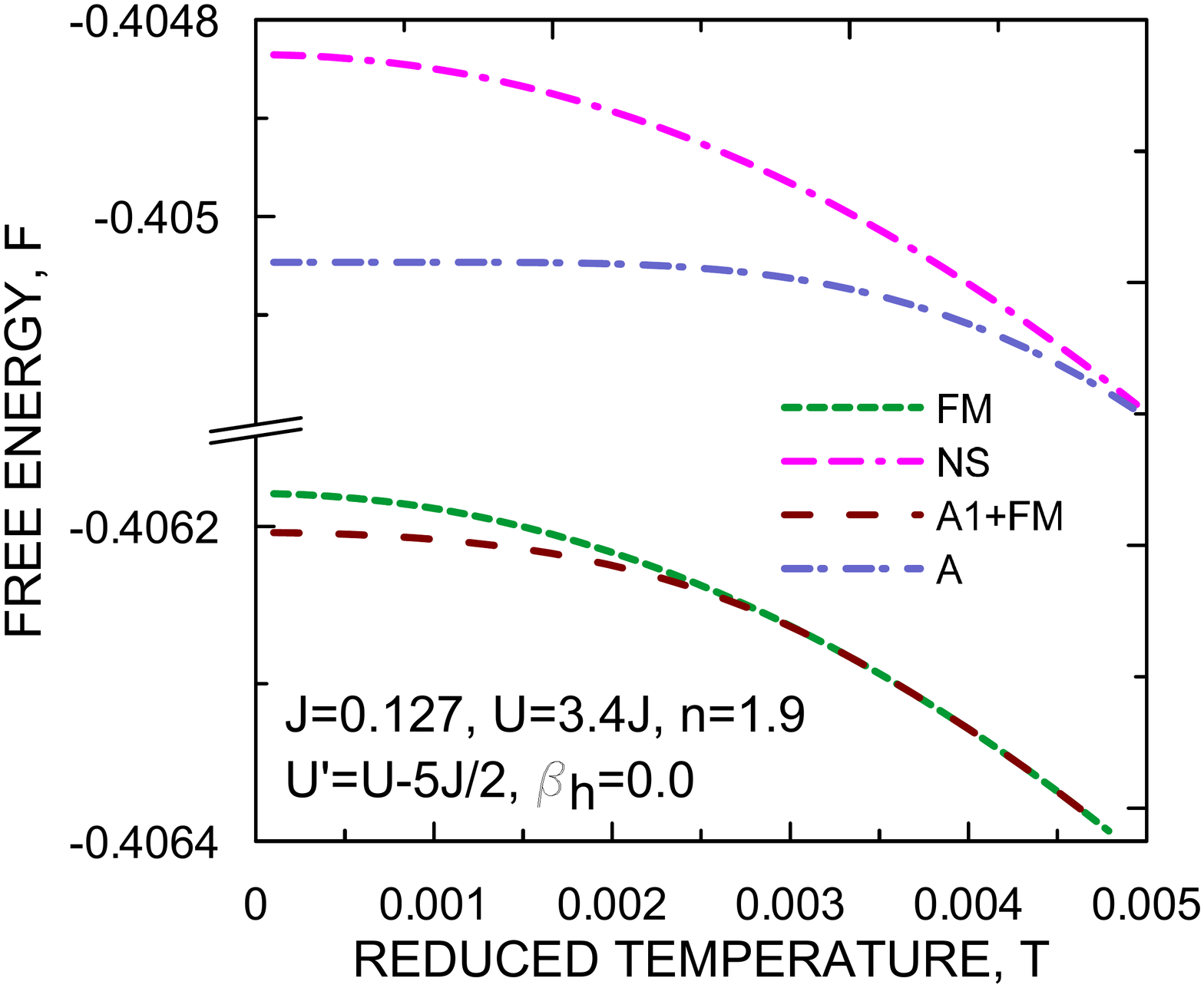} &\\
\end{tabular}
\caption{Color online) (a) Temperature dependence of the free energy for
considered phases, for $n=1.9$ and $J=0.175$ when the SC+AF phase is stable at
$T=0$.
The free-energy values of A and NS phases are very
close, so we exhibit their temperature dependence blown up in part (b).}
\label{fig:fe_SCAF}
\end{figure}

\begin{figure}[htpb]
\centering
\begin{tabular}{ccc}
  (a) & & \quad \quad \quad \quad \quad\\
      & \includegraphics[angle=-90,width=0.35\textwidth]{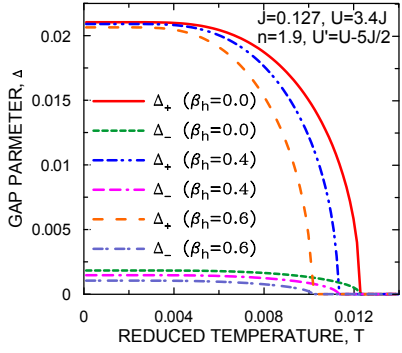} & \\
  (b) & & \quad \quad \quad \quad \quad\\
      & \includegraphics[angle=-90,width=0.35\textwidth]{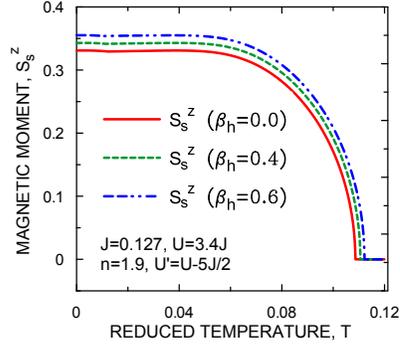} &\\
\end{tabular}
\caption{(Color online) Temperature dependences of the superconducting gaps
$\Delta_+$,
$\Delta_-$ and of the staggered magnetic moment for $n=1.9$, $J=0.175$ and for
selected values of the $\beta_h$ parameter. Note that $T_S<<T_N$.}
\label{fig:D_S_SCAF}
\end{figure}

\begin{figure}[htpb]
\includegraphics[angle=-90,width=0.35\textwidth]{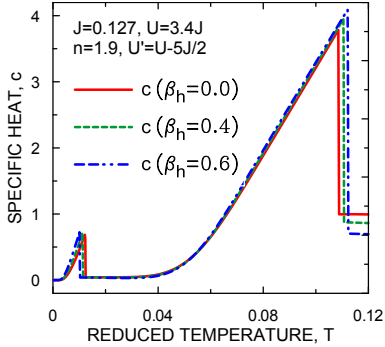} 
\caption{(Color online) Temperature dependence of the specific heat for $n=1.9$,
$J=0.175$ and for
selected values of $\beta_h$
parameter. The behavior is almost independent of $\beta_h$ value and the ratio
$T_N/T_S \approx 10$.}
\label{fig:c_SCAF}
\end{figure}

In Fig. \ref{fig:fe_SCAF} we show temperature dependence of the free energy
for the six considered phases, for the set of microscopic parameters selected
to
make the SC+AF phase stable at $T=0$ and for $\beta_h=0$. Because the
free-energy values of the A and NS phases are very
close, we exhibit their temperature dependences zoomed in Fig.
\ref{fig:fe_SCAF}b. The same is done for the free energy of phases A1+FM and
FM. For the
same values of $n$, $J$, $U$ and $U^{\prime}$, the temperature dependence of the
superconducting
gaps and the staggered magnetic moment in SC+AF phase are shown in Fig.
\ref{fig:D_S_SCAF}, for selected $\beta_h$ values. For given
$\beta_h$ below the superconducting critical temperature $T_S$, the staggered
magnetic
moment and the superconducting gaps have all nonzero values which means, that
we are dealing with the coexistence of superconductivity and antiferromagnetism
in this range of temperatures. Both $\Delta_+$ and
$\Delta_-$ vanish at $T_S$, while the staggered magnetic
moment vanishes at the N\'{e}el temperature, $T_N>>T_S$. In Fig.
\ref{fig:c_SCAF} one
can observe that
there are two typical mean-field discontinuities in the specific-heat at $T_S$
and $T_N$ for a given $\beta_h$. The first of them, at
$T_S$, corresponds to the phase transition from the SC+AF phase to the pure AF 
phase, while the second, at $T_N$, corresponds to the transition from the AF
phase to the
NS phase.  The values of the ratios of the specific heat jump
($\Delta c/c_N$) at $T_C$ that correspond to $\beta_h=0.0$, $0.4$, $0.6$, are
15.075, 16.298, 17.375 respectively. No
antiferromagnetic gap is created, since we have number of electrons $n<2$. The
specific heat discontinuity at AF transition is due to the change of spin
entropy near $T_N$. For $n=2$ the formation of the Slater gap at $T_N$ makes the
superconducting transition to disappear.  
As one can see form Figs. \ref{fig:D_S_SCAF} and \ref{fig:c_SCAF}, with the
increase
of $\beta_h$ the critical temperature $T_S$
is decreasing slightly while the N\'{e}el temperature increases, but the ratio
remains almost fixed, $T_N/T_C\approx 10$. 

\begin{figure}[htpb]
\centering
\begin{tabular}{ccc}
  (a) & & \quad \quad \quad \quad \quad\\
      & \includegraphics[angle=-90,width=0.35\textwidth]{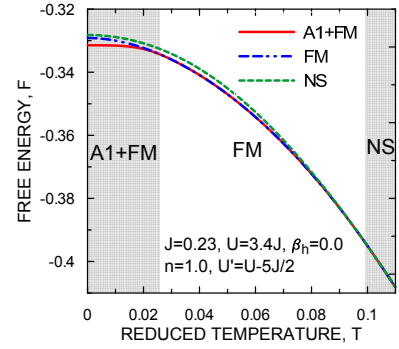} & \\
  (b) & & \quad \quad \quad \quad \quad\\
      & \includegraphics[angle=-90,width=0.35\textwidth]{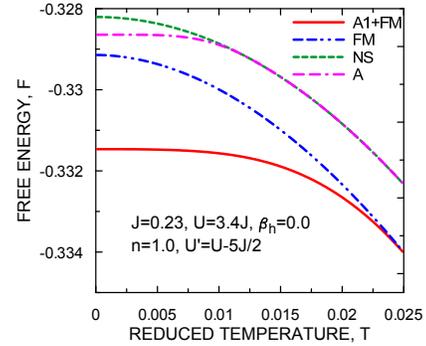} &\\
\end{tabular}
\caption{(Color online) Temperature dependence of the free energy for
$n=1.0$ and $J=0.31625$ when the A1+FM phase is stable at $T=0$. AF
phases does not appear in this case. Free energies for A and A1+FM phases
are shown in
the low-$T$ regime (b) for the sake of clarity.}
\label{fig:fe_A1FM}
\end{figure}

\begin{figure}[htpb]
\centering
\begin{tabular}{ccc}
  (a) & & \quad \quad \quad \quad \quad\\
      & \includegraphics[angle=-90,width=0.35\textwidth]{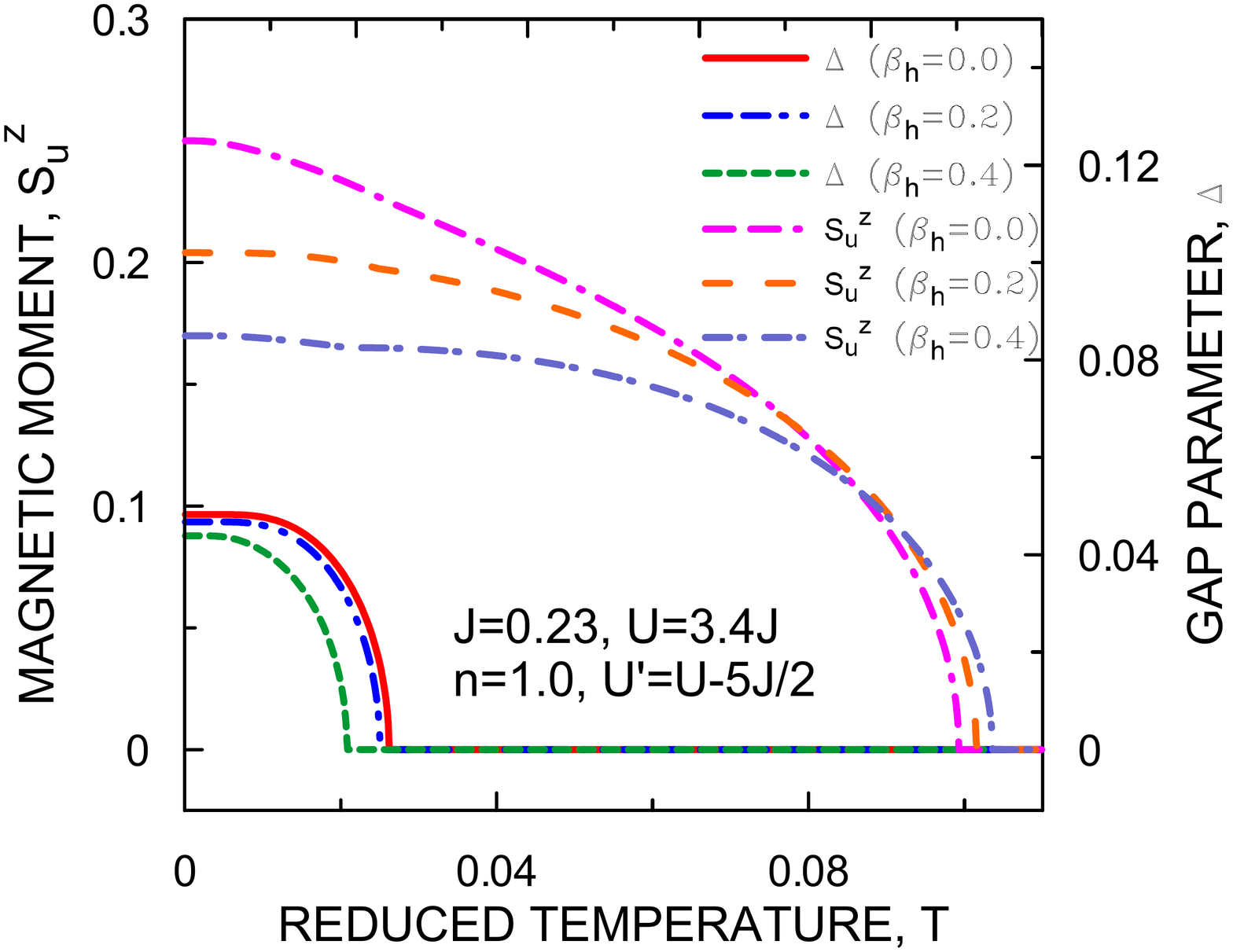} & \\
  (b) & & \quad \quad \quad \quad \quad\\
      & \includegraphics[angle=-90,width=0.35\textwidth]{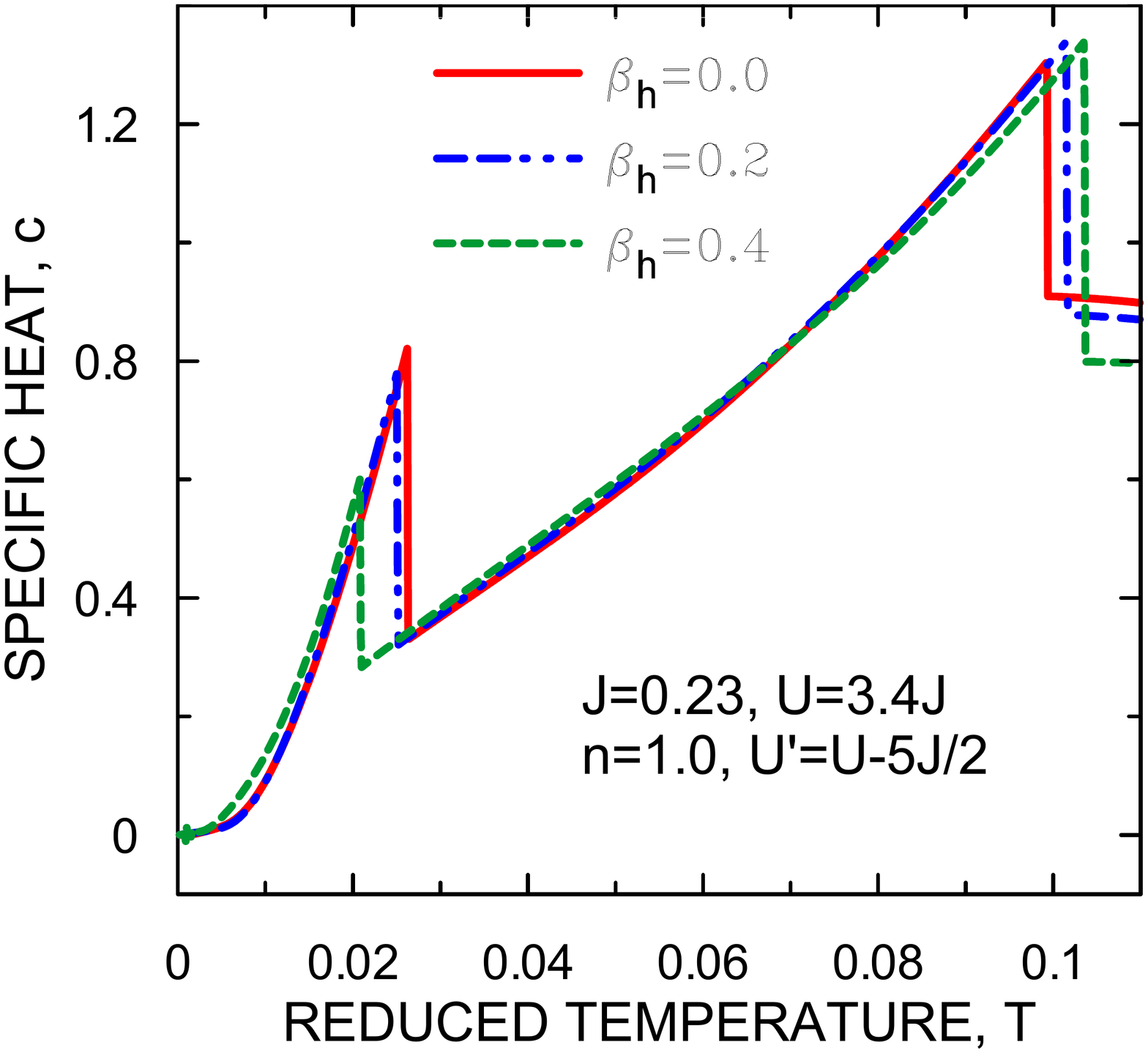} &\\
\end{tabular}
\caption{(Color online) Temperature dependence of the superconducting gaps
$\Delta_+$, $\Delta_-$, magnetic moment (a), and specific heat (b),
for $n=1.0$, $J=0.31625$ and for selected values of $\beta_h$. Qualitative features
do not alter appreciably
even
for $\beta_h=0.4$. The ratio $T_C/T_S\approx 5$.}
\label{fig:D_S_A1FM}
\end{figure}

Temperature dependence of free energies of relevant
phases are presented in Fig. \ref{fig:fe_A1FM} ($\beta_h=0$) for the
microscopic parameters selected to make the A1+FM phase stable at $T=0$.
Free energies for A and A1+FM phases are drawn only in the low-$T$ regime (Fig.
\ref{fig:fe_A1FM}b) for the sake of clarity.
The corresponding temperature dependence
of the superconducting gaps, magnetic moment, and specific heat in A1+FM phase for three
selected values of $\beta_h$ are
shown in Fig. \ref{fig:D_S_A1FM}. Analogously as in the SC+AF case, the
system
undergoes two
phase transitions. The
influence of hybridization on the temperature dependences is also similar to
that in the
case of
coexistence of superconductivity with antiferromagnetism. With the
increasing $\beta_h$, the critical temperature-$T_S$ is
decreasing slightly, whereas the Curie temperature, $T_C$ is slightly
increasing, but still $T_C/T_S\approx 5$. The values of the ratios of the
specific heat jump ($\Delta c/c_N$) at $T_C$ that correspond to $\beta_h=0.0$,
$0.2$, $0.4$, are 1.329, 1.421, 0.793 respectively. 

\begin{figure}[htpb]
\includegraphics[angle=-90,width=0.35\textwidth]{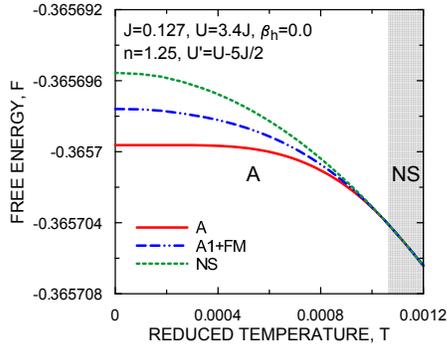} 
\caption{(Color online) Temperature dependences of the superconducting gaps $\Delta_+$,
$\Delta_-$ (a), and the specific heat (b) for $n=1.25$, $J=0.175$ and for
selected
values of $\beta_h$ parameter.}
\label{fig:D_c_A}
\end{figure}

For the sake of completeness, in Fig. \ref{fig:D_c_A}  we provide the temperature
dependence of superconducting gap for the values of parameters that
correspond to 
stable pure superconducting phase of type A at $T=0$ and for three different values of $\beta_h$. In this case, neither the
antiferromagnetically ordered nor the pure ferromagnetic phases
do exist.
As in previous cases, the increasing hybridization decreases $T_S$. It
should be noted that the
values of $\beta_h$ are very close to zero. This is necessary to assume for the
A phase to appear. The values of the ratios of the
specific heat jump ($\Delta c/c_N$) at $T_C$ that correspond to $\beta_h=0.0$,
$0.035$, $0.006$, are 1.382, 1.326, 1.202 respectively.

In Table \ref{table: values_tab} we have assembled the exemplary values of mean field
parameters, chemical potential, as well as free energy for two
different sets of values of microscopic parameters corresponding to the
low-temperature stability of two
considered here superconducting phases: SC+AF and A1+FM. For the two sets of values
of $n$ and $J$, the free energy difference between the stable and first unstable phases is of order $10^{-3}$. The values for
the stable phases are underlined.

\begin{table}[h]
\centering
\caption{Exemplary values of the mean field parameters, the chemical potential,
and the free energy
of the considered phases at $T=10^{-4}$, for two different sets of values of
microscopic parameters: $n$, $J$. The
underlined values correspond the stable phases. The numerical accuracy is better than the last digit.}
 \begin{tabular}{cccc}
  \hline\hline
             &       &  $n=1.9$ & $n=1.0$ \\ 
   parameter & phase &  $J=0.175$ & $J=0.31625$ \\
  \hline
  $\Delta$   &   A   & 0.0097911 & 0.0208481 \\
  $\Delta$   & A1+FM & 0.0056821 & \underline{0.0482677} \\
  $\Delta_+$ & SC+AF & \underline{0.0210081} &     -      \\
  $\Delta_-$ & SC+AF & \underline{0.0017366} &     -     \\
  \hline
  $S^z_u$    &A1+(S)FM& 0.1134254 & \underline{0.2500000}  \\
  $S^z_u$    & (S)FM  & 0.1144301 & 0.2500000  \\
  $S^z_s$    &  SC+AF & \underline{0.3340563} &     -     \\
  $S^z_s$    &   AF   & 0.3314687 &     -     \\
  \hline
  $\mu$     &    A    & -0.0107669& -0.1815757 \\
  $\mu$     &   NS    &-0.0094009 & -0.1799612 \\
  $\mu$     &A1+(S)FM &-0.0175982 & \underline{-0.2530000} \\
  $\mu$     &  (S)FM  &-0.0178066 & -0.253000   \\
  $\mu$     &  SC+AF  &\underline{-0.1708890} & -           \\
  $\mu$     &   AF    &-0.1859011 & -          \\
  \hline
  $F$       &    A    &-0.4050464 & -0.3286443 \\
  $F$       &   NS    &-0.4048522 & -0.3282064 \\
  $F$       &A1+(S)FM &-0.4062039 & \underline{-0.3314652}\\
  $F$       &  (S)FM  &-0.4061793 & -0.3291425 \\
  $F$       &  SC+AF  &\underline{-0.4489338} & -          \\
  $F$       &   AF    &-0.4469097 & -         \\
  \hline\hline
 \end{tabular}
 \label{table: values_tab}
\end{table}


\section{Conclusions and Outlook}\label{sec:conclusions}
We have carried out the Hartree-Fock-BCS analysis of the hybridized two-band
Hubbard model with the Hund's-rule induced magnetism and spin-triplet pairing.
We
have determined the regions of stability of the spin-triplet paired phases with
$\Delta_0\equiv 0$, coexisting with either ferromagnetism (A1+FM) or
antiferromagnetism (SC+AF), as well as pure paired phase (A). We have analyzed in detail
the effect of
inter band hybridization on stability of the those phases.
The hybridization reduces
significantly
the stability regime of the superconducting phase A, mainly in favor of the
paramagnetic (normal) phase, NS. For large enough value of
$\beta_h$ ($\beta_h>0.08$), the A phase disappears altogether. When
it comes to magnetism, with the increase of $\beta_h$, the stability
regime of the saturated ferromagnetically ordered phase is reduced in favor
of the non-saturated. The
influence of the hybridization on the low-temperature stability of the SC+AF
phase is not significant. When the system is close to the half filling, the
SC+AF
phase is the stable one. However, for the half filled band case ($n=2$), the
superconductivity disappears and
only pure antiferromagnetic state survives, since the nesting effect of the two-dimentional band structure prevails then. 

We have also examined the temperature
dependence of the order parameters and the specific heat. For both coexistent
superconducting and magnetically ordered phases (SC+AF and A1+FM) one observes
two separate phase transitions with the increasing temperature. The first of
them, at substantially lower temperature ($T_S$), is the transition from the
superconducting-magnetic coexistent phase to the
pure magnetic phase and the second, occuring at much higher temperature ($T_N
$ or $T_C$), is from the magnetic to the paramagnetic phase (NS). The
hybridization has a negative
influence on the spin-triplet superconductivity, since it reduces the
critical
temperature for each type of the spin-triplet superconducting phase considered
here.
On the other hand, the Curie ($T_C$) and the N\'{e}el ($T_N$) temperatures are
increasing
with the increase of the $\beta_h$ parameter, as it generally increases the
density of states at the Fermi level (for appropriate band fillings). 

One sholud note that since the pairing is intra-atomic in nature the spin-triplet gaps  $\Delta_m$ are of the $s$ type. This constitutes one of the differences with the corresponding situation for superfluid $^3He$, where they are of $p$ type \cite{Anderson}.

It is also important to note that the paired state appears both below and above
the Stoner threshold for the onset of ferromagnetism (cf. Fig.
\ref{fig:diag_stoner}), though its nature changes (A and A1 states,
respectively). In the ferromagnetically ordered phase only the spin-majority
carriers are paired. This is not the case for AF+SC phase. It would be very
interesting to try to detect such highly unconventional SC phase. In particular
the Andreev reflection and in general, the NS/SC conductance spectroscopy will
have an unusual character. We should see progress along this line of research
soon.

As mentioned before all the results presented in the previous
section has been obtained assuming that $U=2.2J$ and $U^{\prime}=U-2J$.
Having said that the value of $J^H=J-U^{\prime}$ determines the strength of
the pairing mechanism while $I=U+J$ is the effective magnetic coupling constant,
one can roughly predict how will the change in the relations between $U$,
$U^{\prime}$, and $J$ result. It seems reasonable to say that the larger is $J$
with respect to $U^{\prime}$ the stronger the superconducting gap in
the paired phases. This would also result in the increase of $T_C$ and a corresponding enlargement
of the area occupied by the superconducting phases on the diagrams. Furthermore
the increase of $U$ with respect to $J$ should result in the increase of the ratios
$T_C/T_S$ and $T_N/T_S$. This is because in that manner we make the magnetic
coupling stronger with respect to the pairing. If we however increase $U$ but do not change $J^H$, then the strength of the pairing would be the
same but the magnetic coupling constant would be stronger so this would favor the
coexistent magnetic and superconducting phases with respect to the pure
superconducting phase. Quite stringent necessary condition for the pairing to
appear $J>U^{\prime}$ (equivalent to $3J>U$ if we assume $U^{\prime}=U-2J$, as
has been done here) indicates that only in specific materials one would expect
for the Hund's rule to create the superconducting phase. This may explain why
only in very few compounds the coexistent ferromagnetic and superconducting phase
has been indeed observed. Obviously, one still has to add the paramagnon pairing (cf. Appendix C).

It should be noted that more exotic magnetic phases may appear in the two-band model \cite{Inagaki1973}. Here, we neglect those phases because of two reasons. First the lattice selected for analysis is bipartite, with strong nesting (AF tendency). Second the additional ferrimagnetic, spiral, etc., phase might appear if we assumed that the second hopping integral $t^{\prime}\neq 0$. Inclusion of $t^{\prime}$ would require a separate analysis, as the lattice becomes frustrated then.

\section{Acknowledgments}\label{sec:ack}
M.Z. has been partly supported by the EU Human Capital Operation Program, Polish
Project No. POKL.04.0101-00-434/08-00. J.S. acknowledges the financial support
from the Foundation for Polish Science (FNP) within project TEAM and
also, the support from the Ministry of Science and Higher Education, through
Grant No. N N 202 128 736.

\section*{Appendix A. Hamiltonian matrix form in the
coexistent SC+AF phase and quasiparticle operators}
In this Appendix we show the general form of the Hamiltonian matrix
$\mathbf{H}_{\mathbf{k}}$ and the pairing operators expressed in terms of the
quasi-particle creation operators from the first step of the diagonalization
procedure disscused in section 2.

For the case of nonzero gap parameters $\Delta_{0 A(B)}$ we have to use eight
element composite creation operator
\begin{equation}
\begin{split}
\mathbf{\tilde{f}}^{\dagger}_{\mathbf{k}}\equiv(\tilde{a}^{\dagger}_{
\mathbf{k}1\uparrow
A},\tilde{a}^{\dagger}_{\mathbf{k}1\downarrow
A},\tilde{a}_{-\mathbf{k}2\uparrow A},\tilde{a}_{-\mathbf{k}2\downarrow
A}, \tilde{a}^{\dagger}_{\mathbf{k}1\uparrow
B},\\
\tilde{a}^{\dagger}_{\mathbf{k}1\downarrow
B},\tilde{a}_{-\mathbf{k}2\uparrow B},\tilde{a}_{-\mathbf{k}2\downarrow
B}),
\nonumber
\end{split}
\end{equation}
to write down the Hamiltonian
(\ref{eq:H_HF}) in the matrix form
\begin{equation}
\hat{H}_{HF}-\mu\hat{N}=\sum_{\mathbf{k}}
\mathbf{\tilde{f}}_{\mathbf{k}}^{\dagger}\mathbf{H}_{\mathbf{k}}
\mathbf{\tilde{f}}_{\mathbf{k}}+2\sum_{\mathbf{k}}(\tilde{\epsilon}_{
\mathbf{k}2 A}+\tilde{\epsilon}_{\mathbf{k}2 B} ) -2\mu N+C,
\label{eq:H_HF_matrixapp}
\end{equation}
where
$\mathbf{\tilde{f}}_{\mathbf{k}}\equiv(\mathbf{\tilde{f}}^{\dagger}_{\mathbf{k}}
)^ { \dagger}$ , and
\begin{widetext}
\begin{equation}
\mathbf{H}_{\mathbf{k}}=\left(\begin{array}{cccccccc}
\tilde{\epsilon}_{\mathbf{k}1A}-\mu & 0 & \delta_{1\mathbf{k}\uparrow\uparrow} &
\delta_{1\mathbf{k}\uparrow\downarrow} & 0 & 0 &
\delta_{3\mathbf{k}\uparrow\uparrow} & \delta_{3\mathbf{k}\uparrow\downarrow}\\
0 & \tilde{\epsilon}_{\mathbf{k}1A}-\mu & \delta_{1\mathbf{k}\downarrow\uparrow}
&
\delta_{1\mathbf{k}\downarrow\downarrow} & 0 & 0 &
\delta_{3\mathbf{k}\downarrow\uparrow} &
\delta_{3\mathbf{k}\downarrow\downarrow}\\
\delta^*_{1\mathbf{k}\uparrow\uparrow} &
\delta^*_{1\mathbf{k}\downarrow\uparrow} &
-\tilde{\epsilon}_{\mathbf{k}2A}+\mu & 0 & \delta_{4\mathbf{k}\uparrow\uparrow}
&
\delta_{4\mathbf{k}\downarrow\uparrow} & 0 & 0\\
\delta^*_{1\mathbf{k}\uparrow\downarrow} &
\delta^*_{1\mathbf{k}\downarrow\downarrow} & 0 &
-\tilde{\epsilon}_{\mathbf{k}2A}+\mu & \delta_{4\mathbf{k}\uparrow\downarrow} &
\delta_{4\mathbf{k}\downarrow\downarrow} & 0 & 0 \\
0 & 0 & \delta^*_{4\mathbf{k}\uparrow\uparrow} &
\delta^*_{4\mathbf{k}\uparrow\downarrow} &
 \tilde{\epsilon}_{\mathbf{k}1B}-\mu & 0 & \delta_{2\mathbf{k}\uparrow\uparrow}
& \delta_{2\mathbf{k}\uparrow\downarrow} \\
0 & 0 & \delta^*_{4\mathbf{k}\downarrow\uparrow} &
\delta^*_{4\mathbf{k}\downarrow\downarrow} & 0
& \tilde{\epsilon}_{\mathbf{k}1B}-\mu & \delta_{2\mathbf{k}\downarrow\uparrow} &
\delta_{2\mathbf{k}\downarrow\downarrow} \\
\delta^*_{3\mathbf{k}\uparrow\uparrow} &
\delta^*_{3\mathbf{k}\downarrow\uparrow} & 0 & 0 &
\delta^*_{2\mathbf{k}\uparrow\uparrow} &
\delta^*_{2\mathbf{k}\downarrow\uparrow} &
-\tilde{\epsilon}_{\mathbf{k}2B}+\mu & 0\\
\delta^*_{3\mathbf{k}\uparrow\downarrow} &
\delta^*_{3\mathbf{k}\downarrow\downarrow} & 0 &
0 & \delta^*_{2\mathbf{k}\uparrow\downarrow} &
\delta^*_{2\mathbf{k}\downarrow\downarrow} & 0 &
-\tilde{\epsilon}_{\mathbf{k}2B}+\mu\\
\end{array} \right).
\label{eq:matrix_Happ2}
\end{equation}
\end{widetext}
The $\delta_{l\mathbf{k}\sigma\sigma^{\prime}}$ are the
generalization of parameters introduced earlier in Eq. (16). 
\begin{equation}
\begin{split}
 \delta_{1\mathbf{k}\sigma\sigma^{\prime}}&=\Delta_{\sigma\sigma^{\prime}
A}U^+_{\mathbf{k}\sigma}U^-_{\mathbf{k}\sigma^{\prime}}+\Delta_{\sigma\sigma^{
\prime}
B}V^+_{\mathbf{k}\sigma}V^-_{\mathbf{k}\sigma^{\prime}},\\
 \delta_{2\mathbf{k}\sigma\sigma^{\prime}}&=\Delta_{\sigma\sigma^{\prime}
A}V^+_{\mathbf{k}\sigma}V^-_{\mathbf{k}\sigma^{\prime}}+\Delta_{\sigma\sigma^{
\prime}
B}U^+_{\mathbf{k}\sigma}U^-_{\mathbf{k}\sigma^{\prime}},\\
 \delta_{3\mathbf{k}\sigma\sigma^{\prime}}&=-\Delta_{\sigma\sigma^{\prime}
A}U^+_{\mathbf{k}\sigma}V^-_{\mathbf{k}\sigma^{\prime}}+\Delta_{\sigma\sigma^{
\prime}
B}V^+_{\mathbf{k}\sigma}U^-_{\mathbf{k}\sigma^{\prime}},\\
 \delta_{4\mathbf{k}\sigma\sigma^{\prime}}&=-\Delta_{\sigma\sigma^{\prime}
A}V^+_{\mathbf{k}\sigma}U^-_{\mathbf{k}\sigma^{\prime}}+\Delta_{\sigma\sigma^{
\prime}
B}U^+_{\mathbf{k}\sigma}V^-_{\mathbf{k}\sigma^{\prime}},
\end{split}
\end{equation}
where $\Delta_{\uparrow\uparrow A(B)}=\Delta_{+1A(B)}$,
$\Delta_{\downarrow\downarrow A(B)}=\Delta_{-1A(B)}$,
$\Delta_{\downarrow\uparrow A(B)}=\Delta_{\uparrow\downarrow
A(B)}=\Delta_{0A(B)}$. 

Below we present the pairing operators expressed in terms of the
quasi-particle creation operators that we have introduced during the first step
of the diagonalization
procedure of the Hamiltonian (\ref{eq:H_HF}).
\begin{equation}
\begin{split}
 \hat{A}^{\dagger}_{\mathbf{k}\sigma
A}&=U^+_{\mathbf{k}\sigma}U^-_{\mathbf{k}\sigma}\tilde{a}^{\dagger}_{\mathbf{k}
1\sigma A}\tilde{a}^{\dagger}_{-\mathbf{k}2\sigma
A}+V^+_{\mathbf{k}\sigma}V^-_{\mathbf{k}\sigma}\tilde{a}^{\dagger}_{\mathbf{k}
1\sigma B}\tilde{a}^{\dagger}_{-\mathbf{k}2\sigma
B}\\
&-U^+_{\mathbf{k}\sigma}V^-_{\mathbf{k}\sigma}\tilde{a}^{\dagger}_{\mathbf{k}
1\sigma A}\tilde{a}^{\dagger}_{-\mathbf{k}2\sigma
B}-V^+_{\mathbf{k}\sigma}U^-_{\mathbf{k}\sigma}\tilde{a}^{\dagger}_{\mathbf{k}
1\sigma B}\tilde{a}^{\dagger}_{-\mathbf{k}2\sigma A},\\
\hat{A}^{\dagger}_{\mathbf{k}\sigma
B}&=U^+_{\mathbf{k}\sigma}U^-_{\mathbf{k}\sigma}\tilde{a}^{\dagger}_{\mathbf{k}
1\sigma B}\tilde{a}^{\dagger}_{-\mathbf{k}2\sigma
B}+V^+_{\mathbf{k}\sigma}V^-_{\mathbf{k}\sigma}\tilde{a}^{\dagger}_{\mathbf{k}
1\sigma A}\tilde{a}^{\dagger}_{-\mathbf{k}2\sigma
A}\\
&+U^+_{\mathbf{k}\sigma}V^-_{\mathbf{k}\sigma}\tilde{a}^{\dagger}_{\mathbf{k}
1\sigma B}\tilde{a}^{\dagger}_{-\mathbf{k}2\sigma
A}+V^+_{\mathbf{k}\sigma}U^-_{\mathbf{k}\sigma}\tilde{a}^{\dagger}_{\mathbf{k}
1\sigma A}\tilde{a}^{\dagger}_{-\mathbf{k}2\sigma B},
\end{split}
\end{equation}
\begin{equation}
\begin{split}
\hat{A}^{\dagger}_{\mathbf{k}0
A}&=\frac{1}{\sqrt{2}}\sum_{\sigma}(U^+_{\mathbf{k}\sigma}U^-_{\mathbf{k}
\bar{\sigma} } \tilde { a } ^ { \dagger } _ { \mathbf { k }
1\sigma A}\tilde{a}^{\dagger}_{-\mathbf{k}2\bar{\sigma}
A}\\
&+V^+_{\mathbf{k}\sigma}V^-_{\mathbf{k}\bar{\sigma}}\tilde{a}^{\dagger}_{
\mathbf { k }
1\sigma B}\tilde{a}^{\dagger}_{-\mathbf{k}2\bar{\sigma}
B}-V^-_{\mathbf{k}\sigma}U^+_{\mathbf{k}\bar{\sigma}}\tilde{a}^{\dagger}_{
\mathbf { k }
1\sigma B}\tilde{a}^{\dagger}_{-\mathbf{k}2\bar{\sigma}
A}\\
&-U^+_{\mathbf{k}\sigma}V^-_{\mathbf{k}\bar{\sigma}}\tilde{a}^{\dagger}_{
\mathbf { k }
1\sigma A}\tilde{a}^{\dagger}_{-\mathbf{k}2\bar{\sigma} B}),\\
\hat{A}^{\dagger}_{\mathbf{k}0
B}&=\frac{1}{\sqrt{2}}\sum_{\sigma}(V^+_{\mathbf{k}\sigma}V^-_{\mathbf{k}
\bar{\sigma} } \tilde { a } ^ { \dagger } _ { \mathbf { k }
1\sigma A}\tilde{a}^{\dagger}_{-\mathbf{k}2\bar{\sigma}
A}\\
&+U^+_{\mathbf{k}\sigma}U^-_{\mathbf{k}\bar{\sigma}}\tilde{a}^{\dagger}_{
\mathbf { k }
1\sigma B}\tilde{a}^{\dagger}_{-\mathbf{k}2\bar{\sigma}
B}+U^-_{\mathbf{k}\sigma}V^+_{\mathbf{k}\bar{\sigma}}\tilde{a}^{\dagger}_{
\mathbf { k }
1\sigma B}\tilde{a}^{\dagger}_{-\mathbf{k}2\bar{\sigma}
A}\\
&+V^+_{\mathbf{k}\sigma}U^-_{\mathbf{k}\bar{\sigma}}\tilde{a}^{\dagger}_{
\mathbf { k }
1\sigma A}\tilde{a}^{\dagger}_{-\mathbf{k}2\bar{\sigma} B}).
\end{split}
\end{equation}


\section*{Appendix B. Hamiltonian matrix and quasiparticle states for the coexistent ferromagnetic-spin-triplet superconducting phase}
In this Appendix we show briefly the approach to the coexistent
ferromagnetic-spin-triplet superconducting phase within the mean-field-BCS
approximation. Inanalogy to the situation considered in Section 2, we make use
of relations (\ref{eq:Hund_pairing}), (\ref{eq:U_prime}) and transform our
Hamiltonian into the reciprocal space to get
\begin{equation}
\begin{split}
\hat{H}_{HF}-\mu\hat{N}&=\sum_{\underline{\mathbf{k}}l\sigma}(\epsilon_{
\mathbf { k}}-\mu-\sigma
IS^z_u)\hat{n}_{\mathbf{k}l\sigma}\\
&+\sum_{\mathbf{k}
ll^ { \prime}(l\neq l^{\prime})\sigma}\epsilon_{12\mathbf{k}}
a_{\mathbf{k}l\sigma}^{\dag}a_{\mathbf{k}l^{\prime}
\sigma} \\
&+\sum_{\mathbf{k},m=\pm
1}(\Delta_{m}^{*}\hat{A}_{\mathbf{k},m}+\Delta_{m}\hat{A}_{
\mathbf{k},m}^{\dagger})\\
&+\sqrt{2}\sum_{\mathbf{k}}
(\Delta_{0}^{*}\hat{A}_{
\mathbf{k},0}+\Delta_{0}\hat{A}_{\mathbf{k},0}^{\dagger}) \\
&+N\bigg\{\frac{|\Delta_1|^2+|\Delta_{-1}|^2+2|\Delta_{0}
|^2}{J-U^{\prime}}+2I(S^z_u)^2\bigg\},
\end{split}
\label{eq:h_HF_calosc_A1FM}
\end{equation}
where $S^z_u$ is the uniform average magnetic moment and this time the sums are
taken over all N independent $\mathbf{k}$ points, as here we do not need to
perform the division into two sublattices. In the equation above we have
omitted the terms that only lead to the shift of the
reference energy. Next, we diagonalize the one
particle part of the H-F Hamiltonian by introducing quaziparticle operators 
\begin{equation}
\begin{split}
\tilde{a}_{\mathbf{k}1\sigma}=\frac{1}{\sqrt{2}}(a_{
\mathbf{k}1\sigma}+a_{\mathbf{k}2\sigma}),\\
\tilde{a}_{\mathbf{k}2\sigma}=\frac{1}{\sqrt{2}}(-a_{
\mathbf{k}1\sigma}+a_{\mathbf{k}2\sigma}),
\label{eq:kwaziczastki}
\end{split}
\end{equation}
with dispersion relations
\begin{equation}
\begin{split}
\tilde{\epsilon}_{\mathbf{k}1\sigma}&=\epsilon_{\mathbf{k}
}-\mu- \sigma IS^z + |\epsilon_{12\mathbf{k}}|,\\
\tilde{\epsilon}_{\mathbf{k}2\sigma}&=\epsilon_{\mathbf{k
}}-\mu- \sigma IS^z - |\epsilon_{12\mathbf{k}}|.
\label{eq:eigenvalues_S_22}
\end{split}
\end{equation}
Using the 4-component composite creation operator 
$\mathbf{\tilde{f}}^{\dagger}_{\mathbf{k}}\equiv(\tilde{a}^{\dagger}
_{\mathbf{k}1\uparrow},\tilde{a}^{\dagger}_{\mathbf{k}
1\downarrow},\tilde{a}_{-\mathbf{k}2\uparrow},\tilde{a}_{-\mathbf{k}2\downarrow
})$, we can construct the 4x4 Hamiltonian matrix and write it in the following form
\begin{equation}
\hat{H}_{HF}-\mu\hat{N}=\sum_{\mathbf{k}}
\tilde{\mathbf{f}}_{\mathbf{k}}^{\dagger}\mathbf{\tilde{H}}_{
\mathbf{k}}\tilde{\mathbf{f}}_{\mathbf{k}}+\sum_{
\mathbf{k}\sigma}\tilde{\epsilon}_{\mathbf{k}2\sigma}+C,
\end{equation}
where
\begin{equation}
\mathbf{\tilde{H}}_{\mathbf{k}}=\left(\begin{array}{cccc}
\tilde{\epsilon}_{\mathbf{k}1\uparrow} & 0 & \Delta_1 & \Delta_{0}\\
0 & \tilde{\epsilon}_{\mathbf{k}1\downarrow} & \Delta_{0} &
\Delta_{-1}\\
\Delta_1^* & \Delta_0^* & -\tilde{\epsilon}_{\mathbf{k}2\uparrow} &
0\\
\Delta_0^* & \Delta_{-1}^* & 0 &
-\tilde{\epsilon}_{\mathbf{k}2\downarrow}
\end{array} \right),
\label{eq:matrix_H}
\end{equation}
with
$\mathbf{\tilde{f}}_{\mathbf{k}}\equiv(\mathbf{\tilde{f}}^{\dagger}_
{\mathbf{k}})^{\dagger}$. Symbol $C$ refers to the last two terms of r. h.
s. of expression (\ref{eq:h_HF_calosc_A1FM}).
After making the diagonalization transformation of (\ref{eq:matrix_H}) we can
write the H-F Hamiltonian as follows
\begin{equation}
\hat{H}_{HF}-\mu\hat{N}=\sum_{\mathbf{k}l\sigma}\lambda_{\mathbf{k}
l\sigma}\alpha^{\dagger}_{\mathbf{k}l\sigma}\alpha_{
\mathbf{k}l\sigma}+\sum_{\mathbf{k}\sigma}(\tilde{\epsilon}_{
\mathbf{k}2\sigma}-\lambda_{\mathbf{k}2\sigma})+C,
\label{eq:hamiltonian_final}
\end{equation}
where we have again introduced the quasiparticle operators
$\alpha_{\mathbf{k}l\sigma}$ and $\alpha^{\dagger}_{\mathbf{k}l\sigma}$.
Assuming that $\Delta_0=0$ and that the remaining gap parameters are real, we
can write down the dispersion relations for the quasi-particles
$\lambda_{\underline{\mathbf{k}}l\sigma}$ in the following way
\begin{equation}
\begin{split}
\lambda_{\mathbf{k}1\uparrow}=\sqrt{(\epsilon_{\mathbf{k
}}-\mu-IS^z)^2+\Delta_1^2}+\beta_h|\epsilon_{\mathbf{k}}|,\\
\lambda_{\mathbf{k}1\downarrow}=\sqrt{(\epsilon_{\mathbf{
k}}-\mu+IS^z)^2+\Delta_{-1}^2}+\beta_h|\epsilon_{\mathbf{k}}|,\\
\lambda_{\mathbf{k}2\uparrow}=\sqrt{(\epsilon_{\mathbf{k}
}-\mu-IS^z)^2+\Delta_1^2}-\beta_h|\epsilon_{\mathbf{k}}|,\\
\lambda_{\mathbf{k}2\downarrow}=\sqrt{(\epsilon_{\mathbf{
k}}-\mu+IS^z)^2+\Delta_{-1}^2}-\beta_h|\epsilon_{\mathbf{k}}|.\\
\end{split}
\label{eq:disp_rel_A_S}
\end{equation}
In this manner we have obtained the fully diagonalized Hamiltonian analytically for the case
of superconductivity coexisting with ferromagnetism. Next, in the similar way as
for the antiferromagnetically ordered phases, we can construct the set of self
consistent equations for the mean field parameters $\Delta_{\pm 1}$, $S^z_u$ and
for the chemical potential, as well as construct the expression for the free energy.

\section*{Appendix C. Beyond the Hartree-Fock approximation: Hubbard-Stratonovich transformation}

In outlining the systematic approach going beyond the Hartree-Fock approximation we start with Hamiltonian (\ref{eq:H_start2}) with
the singlet pairing part $\sim (U^{\prime}+J)\sum_{i}B_i^{\dagger}B_i$
neglected,
i.e.
\begin{equation}
\hat{H}=\hat{H}_0
+ U\sum_{il}\hat{n}_{il\uparrow}\hat{n}_{
il\downarrow}-J^H\sum_ {im}\hat{A}^{\dagger}_{im} \hat{A}_{im},
\label{eq:H_start_ApC}
\end{equation}
where $\hat{H}_0$ contains the hopping term, and $J^H\equiv J-U^{\prime}$. We
use the spin-rotationally invariant form of the Hubbard term
\begin{equation}
\hat{n}_{il\uparrow}\hat{n}_{il\downarrow}=\frac{\hat{n}^2_{il}}{4}-(\vec{\mu}
_{il} \cdotp
\mathbf{\hat{S}}_{il} )^2,
\label{eq:Hub_term}
\end{equation}
where $\hat{n}_{il}=\sum_{\sigma}\hat{n}_{il\sigma}$ and $\vec{\mu}_i$ is an
arbitrary unit vector establishing local spin quantization axis. One should note
that, strictly speaking we have to make the Hubbard-Stratonovich transformation
twice, for each of the last two terms in (\ref{eq:Hub_term}) separately. The
last term will be effectively transformed in the following manner
\begin{equation}
\begin{split}
-J^H\sum_ {im}\hat{A}^{\dagger}_{im}& \hat{A}_{im} \rightarrow \\
&-\sum_{im}(\hat{A}_{im}^{\dagger}\Delta_{im}+\hat{A}_{im}\Delta_{im}^* -
|\Delta_{im}|^2/J^{H}),
\end{split}
\end{equation}
where $\Delta_{im}$ is the classical (Bose) field in the coherent-state representation. The term (\ref{eq:Hub_term}) can be represented in the standard form through the Poisson integral
\begin{equation}
\exp\bigg(\frac{\hat{\alpha}_i^2}{2}\bigg)=\frac{1}{\sqrt{2\pi}}\int_{-\infty}^{\infty}dx_i\bigg(-\frac{x_i^2}{2}+\hat{\alpha}_ix_i\bigg).
\end{equation}
In effect, the partition function for the Hamiltonian (\ref{eq:H_start_ApC})
will have the form in the coherent-state representation
\begin{equation}
\begin{split}
 \mathcal{Z}&=\int \mathcal{D}[a_{il\sigma},a^{\dagger}_{il\sigma}, \Delta_{im},
\Delta^*_{im}, \lambda_{il}] \\
&\times\exp\bigg\{-\int_0^\beta d\tau
\bigg\{\sum_{ijll^{\prime}\sigma}a^{\dagger}_{il\sigma}\bigg[t_{ij}^{ll^{\prime}}
+\bigg(\frac { \partial}{\partial \tau}-\mu
\bigg)\delta_{ij}\delta_{ll^{\prime}} \bigg]a_{jl^{\prime} \sigma}\\
&-\sum_{im}\bigg[\Delta_{im}(\tau)
\hat{A}^{\dagger}_{im}(\tau)+\Delta^*_{im}(\tau)
\hat{A}_{im}(\tau)-\frac{|\Delta_{im}(\tau)|^2}{J^H}\bigg]\\
&-\sum_{il}\sqrt{2}\lambda_{il} \vec{\mu}_{il} \cdotp \mathbf{\hat{S}}_{il} + \lambda_{il}^2 \bigg\}
\bigg\},
\end{split}
\label{eq:grass}
\end{equation}
where we have included only the spin and the pairing fluctuations. In the
present paper
$t_{ij}^{ll^{\prime}}=t_{ij}\delta_{ll^\prime}+(1-\delta_{ll^{\prime}})t_{ij}^{1
2}$. Also the integration takes place in imaginary time domain and the creation
and anihilation operators are now Grassman variables \cite{Negele}. In this
formulation $\Delta_{im}$ and $\lambda_i$ represent local fields which can be
regarded as mean (Hartree-Fock) fields with Gaussian fluctuations.

With the help of (\ref{eq:grass}) we can define "time dependent" effective
Hamiltonian.
\begin{equation}
\begin{split}
 \hat{H}(\tau)&\equiv
\sum_{ijll^{\prime}\sigma}t^{ll^{\prime}}_{ij}a^{\dagger}_{il\sigma}(\tau)a_{
jl^{\prime}\sigma}(\tau)-J^H\sum_{im}\bigg[\Delta_{im}(\tau)
\hat{A}^{\dagger}_{im}(\tau)\\
&+\Delta^*_{im}(\tau)
\hat{A}_{im}(\tau)-|\Delta_{im}(\tau)|^2\bigg]\\
&-U\sum_i\bigg[\vec{\lambda}
_{il}(\tau)\cdotp\mathbf{ \hat{S}}_{il}(\tau)+\frac{\vec{\lambda}_{il}^2(\tau)}{2} \bigg],
\end{split}
\end{equation}
where now the fluctuating dimensionless fields are defined as
\begin{equation}
 \vec{\lambda}_{il}(\tau)\equiv\frac{\sqrt{2}\vec{\mu}_{il}\lambda_{il}(\tau)}{U},
\quad \Delta_{im}(\tau)\equiv\frac{\Delta_{im}(\tau)}{J^H}. 
\end{equation}
Note that the magnetic molecular field $\sim U\vec{\lambda}_{il}(\tau)$ is substantially stronger than the pairing field $\sim J^H \Delta_{im}(\tau)$. In the saddle point approximation
$\vec{\lambda}_{il}(\tau)\equiv \lambda_{il} \mathbf{e}_z$,
$\Delta_{im}(\tau)=\Delta^*_{im}(\tau)\equiv \Delta$, and we obtain the
Hartree-Fock-type approximation. Therefore, the quantum fluctuations are described by
the terms
\begin{equation}
\begin{split}
 -U\sum_{il}& \vec{\delta \lambda}
_{il}(\tau)\cdotp \mathbf{\hat{S}}_{il}(\tau)\\
-&J^H\sum_{im}\bigg[\delta\Delta_{im}(\tau)
\hat{A}^{\dagger}_{im}(\tau)+\delta\Delta^*_{im}(\tau)
\hat{A}_{im}(\tau)\bigg].
\end{split}
\end{equation}
The first term represents the quantum spin fluctuations of the amplitude
$ \vec{\delta \lambda}_{il}(\tau)\equiv \vec{\lambda}_{il}(\tau)-\lambda\mathbf{e}_z$, the second
decribes pairing fluctuations. Both fluctuations are Gaussian due to the presence of
the terms $\sim \vec{\delta\lambda}^2_{il}(\tau)$ and $|\delta\Delta_{im}(\tau)|^2$. In other
words, they represent the higher-order contributions and will be treated in
detail elswhere. In such manner, the mean-field part (real-space pairing) and
the fluctuation part ( pairing in $\mathbf{k}$ space) can be incorporated thus
into a single scheme.

\end{document}